\documentclass[aps,prd,nofootinbib,floats,
floatfix,preprintnumbers,groupedaddress]{revtex4}
\usepackage{comment}
\usepackage{graphicx}
\usepackage{amsmath, mathrsfs}
\usepackage{float}
\usepackage{subfigure}
\usepackage[usenames]{color}
\usepackage{amssymb}
\usepackage{bbm}
\usepackage{csquotes}
\usepackage{color}
\usepackage{psfrag}
\usepackage{graphicx}
\usepackage{orcidlink}
\usepackage{caption,subcaption}
\usepackage{marginnote}
\usepackage[utf8]{inputenc}
\usepackage{tikz}
\usetikzlibrary{positioning,decorations.pathmorphing}

\usepackage{epstopdf}
\newcommand{\bea}{\begin{aligned}}
\newcommand{\eea}{\end{aligned}}
\newcommand{\beq}{\begin{equation}}
\newcommand{\eeq}{\end{equation}}

\newcommand{\bse}{\begin{subequations}}
\newcommand{\ese}{\end{subequations}}

\usepackage{hyperref}
\hypersetup{
     colorlinks   = true,
     citecolor    = red,
     linkcolor    = blue,
     urlcolor     = blue,
}

\newcommand{\bmm}{\begin{multline}}
\newcommand{\emm}{\end{multline}}

\begin{document}
\title{\Large{Black Holes in Lorentz-Violating Gravity: Thermodynamics, Geometry, and Particle Dynamics}}
%%%%%%%%%%%%%%%%%%%
\author{Ankit Anand$^1$\orcidlink{0000-0002-8832-3212}, Aditya Singh$^2$\orcidlink{0000-0002-2719-5608}, and Anshul Mishra$^3$}
\email{Anand@iitk.ac.in} \email{24pr0148@iitism.ac.in} \email{anshulmishra2025@gmail.com}
\affiliation{$^1$Department of Physics, Indian Institute of Technology Kanpur, Kanpur 208016, India.\\
$^2$Department of Physics, Indian Institute of Technology (Indian School of Mines), Dhanbad, Jharkhand-826004, India.\\
$^3$Department of Physics, Indian Institute of Technology Madras, Chennai 600036, India.}
%%%%%%%%%%%%%%%

%%%%%%%%%%%%%%%
\author{Saeed Noori Gashti\orcidlink{0000-0001-7844-2640}}
\email{saeed.noorigashti@stu.umz.ac.ir; sn.gashti@du.ac.ir}
\affiliation{School of Physics, Damghan University, P. O. Box 3671641167, Damghan, Iran.}
%%%%%%%%%%%%%%%
\author{Takol Tangphati$^4$ and Phongpichit Channuie$^{4,5}$}
\email{takol.ta@mail.wu.ac.th} \email{phongpichit.ch@mail.wu.ac.th}
\affiliation{$^4$ School of Science, Walailak University, Nakhon Si Thammarat, 80160, Thailand.\\
$^5$ College of Graduate Studies, Walailak University, Nakhon Si Thammarat, 80160, Thailand.}
%%%%%%%%%%%%%%%%%%%%%
%\author{}
%\\ \textcolor{red}%{\;\;\;\;\;\;\;\;\;\;\;\;\;\;\;\;\;\;\;\;\;\;\;\;\;\\\;\;\;\;\;\;\;\;\;\;\;\;\;%\;\;\;\;\;\;\;\;\;\;\;\; All authors contributed equally and Names are written alphabetically.}
%\email{}
%\affiliation{}
%%%%%%%%%%%%%%%%%%%%%%%%%%

%%%%%%%%%%%%%%%%%%%%%%%%%%
\pagenumbering{arabic}

\renewcommand{\thesection}{\arabic{section}}
%%%%%%%%%%%%%%%%%%%%%
\begin{abstract}
\begin{center}
    \hspace{1cm}
\end{center}

 We investigate the thermodynamics, topology, and geometry of black holes in Lorentz-violating gravity. Modifications in the theory by perturbative parameter lead to coupled changes in horizon structure and thermodynamic behaviour, allowing us to derive generalized universal relations and explore implications for the Weak Gravity Conjecture. The thermodynamic topology reveals distinct topological charges, with photon spheres identified as robust topological defects. Our analysis shows that the Ruppeiner curvature remains universally negative across thermodynamic ensembles, indicating dominant attractive interactions among microstructures. This ensemble-independent behaviour highlights a fundamental thermodynamic universality in Lorentz-violating settings. Together, these results provide a consistent and rich framework for understanding black hole microphysics and gravitational consistency in modified theories. We further study the motion of timelike test particles in these black hole spacetimes by analyzing the effective potential shaped by the Lorentz-violating couplings. The resulting dynamics reveal the existence of bound orbits and stable circular trajectories, with the location of the innermost stable circular orbit and turning points significantly influenced by the parameters $\ell_{1,2}$, and the cosmological constant. Numerical simulations of trajectories in the $x-y,\,x-z$, and 3D planes show precessing, bounded, and plunging orbits, depending on the particle's specific energy and angular momentum. These results highlight how Lorentz-violating effects alter the structure of geodesic motion and provide potential observational signatures in the dynamics of massive particles near black holes.

\end{abstract}

%%%%%%%%%%%%%%%%%%%%%%

\maketitle

%\textbf{Keywords: Kalb-Ramond field, Black holes, Massive particle trajectories}

\section{Introduction}

The observational era of black hole physics was revolutionized with the pioneering results of the Event Horizon Telescope (EHT) Collaboration, which successfully captured the first images of black hole shadows \cite{EventHorizonTelescope:2019dse, EventHorizonTelescope:2019uob, EventHorizonTelescope:2022wkp, EventHorizonTelescope:2021bee, EventHorizonTelescope:2021srq, EventHorizonTelescope:2022urf, EventHorizonTelescope:2022xqj}. These shadow images offer direct access to the strong gravity regime near event horizons, serving as powerful probes of general relativity and potential modifications to it. The concept of black hole shadows, however, is not new; its theoretical roots can be traced back to Synge’s early work on the propagation of light in curved spacetime \cite{Synge:1966okc}, and Bardeen’s seminal analysis of photon orbits in Schwarzschild and Kerr geometries \cite{Bardeen:1972fi}. Astrophysical black holes, however, are rarely isolated. They typically reside in complex environments, such as expanding cosmological backgrounds or accretion-dominated systems. These contexts necessitate a more nuanced understanding of shadow formation. In particular, the presence of magnetized or dispersive media—such as plasma—can influence light trajectories, thereby altering the observed shadow profile. Extensive investigations by Perlick, Tsupko, and collaborators \cite{Perlick:2015vta, Tsupko:2021yca, Bisnovatyi-Kogan:2015dxa, Perlick:2023znh} have elucidated how plasma effects modify both light deflection and shadow morphology. In recent years, black hole shadows have emerged not only as observational targets but also as theoretical tools for testing gravitational physics. Numerous studies have analyzed shadow structures in a wide variety of spacetime geometries, including static, stationary, rotating, and dynamical black holes \cite{Hod:2012ax, Khoo:2016xqv, Decanini:2010fz, Cederbaum:2015fra, Johannsen:2013vgc, Teo:2003ltt, Okyay:2021nnh, Kuang:2022xjp, Ovgun:2024zmt, Uniyal:2023ahv, Pulice:2023dqw, Yang:2023tip, Cimdiker:2023zdi, Kumaran:2023brp, Ovgun:2023ego, Atamurotov:2022knb, Cimdiker:2021cpz}. These efforts underscore the sensitivity of the shadow boundary to a range of physical parameters, including mass, spin, charge, and the surrounding matter content. Moreover, the geometric properties of shadows offer novel cosmological applications. It has been proposed that shadows can serve as standard rulers for distance estimation \cite{Tsupko:2019pzg, Vagnozzi:2020quf} and as diagnostic tools to distinguish between classical and alternative black hole models \cite{Vagnozzi:2022moj}, making them valuable across both observational and theoretical frontiers of gravitational research.

The quantum nature of gravity remains one of the most profound challenges in theoretical physics. While general relativity has been remarkably successful in describing gravitational phenomena on macroscopic scales, it is widely believed to be incomplete at high energies where quantum effects become significant. A range of candidate theories of quantum gravity—including string theory, loop quantum gravity, and the AdS/CFT correspondence—have been proposed to address this shortcoming~\cite{Birrell:1982ix, Maldacena:1997re, Aharony:1999ti, Gubser:1998bc, Alfaro:1999wd, Alfaro:2001rb, Rovelli:1989za}. However, experimental signatures of quantum gravity are typically expected to manifest near the Planck scale, a regime far beyond current technological reach. Interestingly, many of these quantum gravity proposals predict subtle deviations from established symmetries, such as Lorentz invariance, which may be testable at lower energy scales. In particular, spontaneous Lorentz symmetry breaking has emerged as a key feature in several string-inspired models. This has led to the development of effective field theories that incorporate Lorentz-violating terms while remaining consistent with known physics. Among these, the Standard Model Extension  provides a comprehensive framework for systematically studying Lorentz violation across both particle physics and gravity~\cite{Kostelecky:1988zi, Colladay:2001wk, Kostelecky:2000mm, Kostelecky:2001mb}. Bumblebee gravity and those involving Kalb-Ramond fields~ \cite{Do:2020ojg, Liu:2024oas, Duan:2023gng, Yang:2023wtu, Lessa:2019bgi, Kumar:2020hgm, Liu:2024lve} represent compelling avenues for investigating the effects of Lorentz violation. These modified gravity theories allow for spontaneous symmetry breaking through the dynamics of vector or tensor fields acquiring vacuum expectation values. Such mechanisms not only enrich the theoretical landscape but also give rise to novel black hole solutions with distinct phenomenological features. Quantum corrections to black hole thermodynamics have become a central focus in the quest to reconcile gravity with quantum mechanics. These corrections, typically non-perturbative in nature, introduce modifications to the Bekenstein-Hawking entropy, leading to subleading terms that become increasingly relevant in the regime of small black holes \cite{Faulkner:2013ana, Harlow:2014yka, Sen:2012dw}. Such quantum-induced modifications influence not only the entropy but also alter the mass spectrum and thermodynamic stability of black hole solutions. Notably, for Schwarzschild and Schwarzschild-AdS black holes, the corrected entropy leads to a reduction in the effective mass and induces significant changes in stability criteria at small horizon areas \cite{Pourhassan:2020yei, Pourhassan:2024yfg}. The statistical foundation of these corrections has been rigorously examined via quantum-corrected partition functions, enabling a consistent derivation of thermodynamic identities. In particular, the generalised Smarr and Gibbs-Duhem relations have been reformulated to incorporate these quantum contributions, preserving the integrability of the thermodynamic phase space \cite{Dehghani:2008qr}. To probe the universality of the thermodynamic extremality condition, one considers a perturbation to the gravitational action, introduced through a small parameter $\varepsilon$, and scaled relative to the cosmological constant. This deformation induces modifications in both the spacetime geometry and the associated thermodynamic quantities. Within this framework, one arrives at a universal relation characterizing extremal black holes \cite{Goon:2019faz}, given by
\begin{equation}\label{Extremality condition}
\frac{\partial M_{\text{ext}}(\vec{\mathcal{Q}}, \varepsilon)}{\partial \varepsilon} = \lim_{M \to M_{\text{ext}}} -T \left( \frac{\partial S(M, \vec{\mathcal{Q}}, \varepsilon)}{\partial \varepsilon} \right)_{M, \vec{\mathcal{Q}}} \ ,
\end{equation}
where $M_{\text{ext}}$ denotes the extremal mass (defined at zero temperature), $S$ is the black hole entropy, and $\vec{\mathcal{Q}}$ represents the set of conserved thermodynamic charges. This relation reflects how extremal solutions evolve under infinitesimal perturbations and provides a universal constraint on the thermodynamic behavior of gravitational systems near extremality.

A photon sphere \cite{Claudel:2000yi, Shoom:2017ril} refers to a hypersurface comprised entirely of closed null geodesics, representing the critical zone beyond which light is either captured by or escapes from a compact object. It defines the innermost region where light can be gravitationally confined to orbit, and exists in both unstable and stable configurations. The unstable photon sphere is particularly significant in the context of black hole imaging and shadow formation, as minor perturbations result in photons either spiraling inward or escaping to infinity. In contrast, stable photon spheres, while rarer, are often associated with dynamical instabilities in the underlying spacetime. The photon sphere in Schwarzschild geometry restricts the motion to the equatorial plane ($\theta=\pi/2$), and using the condition for the lightlike geodesics, there are two conserved quantities along a geodesic: the energy and angular momentum. The radial equation of motion
\begin{eqnarray}\label{Energy Relation Schwarzschild}
    \left(\frac{dr}{d\lambda}\right)^2 + V(r) = \mathcal{E}^2 \;\;\;\;\;\text{where}\;\;\;\;\; V(r) = \frac{L^2}{r^2}\left(1-\frac{2M}{r}\right) \ , 
\end{eqnarray}
where $\lambda$ is an affine parameter. Circular photon orbits occur where the first and second derivative of $r$ with respect to the affine parameter vanishes. Using Eq.~\eqref{Energy Relation Schwarzschild}, one can easily verify that computing the condition $\partial_rV(r)=0$ yields the radius of photon sphere as $r=3M$. In fact, one finds that for a photon, the only closed orbit is a circular one with $r=3M$. All other photon orbits either fall into the hole or escape. The photon sphere at $r=3M$ represents an unstable last orbit for light. Physically, at this radius, the spacetime curvature is strong enough to bend a light ray into a circular path. However, this balance is unstable: any small radial perturbation causes the photon to spiral away or fall in. Indeed, small deviations from $r=3M$ send the photon inward or outward rather than remaining bound.  In the effective-potential picture, $V(r)$ has a local maximum at $r=3M$, confirming instability. Photons with tangential paths at $r<3M$ inevitably spiral into the black hole, while those at $r>3M$ can escape to infinity. Thus, $r=3M$ is the last possible circular photon orbit, sometimes called the last photon orbit or photon sphere. The photon sphere and its uniqueness teach us a lot about black hole optics and astrophysics \cite{Virbhadra:2022iiy, Virbhadra:1999nm, Virbhadra:1998dy, Cederbaum:2015fra}. It effectively determines the shadow of a black hole and the formation of the bright photon ring seen by distant observers. Light emitted from background sources near the photon sphere can orbit one or more times before escaping, producing multiple lensed images and a bright ring. For example, NASA notes that thin rings of light appear at the edge of the black hole shadow due to photons that loop around the hole. This effect is confirmed by the Event Horizon Telescope image of $M87$: the bright ring in the EHT image arises from light near the photon sphere~\cite{EventHorizonTelescope:2019dse, EventHorizonTelescope:2019uob, EventHorizonTelescope:2019jan, EventHorizonTelescope:2019ths}.

Black holes are the solutions of $G_{\mu\nu}= 8\pi  T_{\mu\nu}$. While other defect-like solutions (e.g., strings, branes) exist, which can be endowed with topological charges to probe their global and local geometric structure. Gibbons and Hawking~\cite{Gibbons:1976ue} introduced a Euclidean path integral approach to black hole thermodynamics, relating the partition function to the classical Euclidean action. However, this formulation initially exhibited instabilities, such as negative specific heat. York~\cite{York:1986it} resolved these by placing the black hole in a thermal cavity, stabilizing the ensemble and defining a consistent thermodynamic description for massive black holes. York treated mass and temperature as independent variables, extending the thermodynamic phase space. This generalized free energy reduces to the standard form when the equilibrium relation between mass and temperature is imposed. Certainly. The generalized free energy functional is defined in terms of the system's energy and entropy, with a free parameter $\tau$ possessing the dimension of time. This parameter is interpreted as the inverse temperature of an external thermal reservoir or cavity. The generalized free energy is generally off-shell and attains its on-shell form only when $\tau$ equals the inverse Hawking temperature, at which point the black hole solution satisfies the Einstein field equations. A thermodynamic vector field is defined by the gradient of the generalized free energy with respect to the horizon radius, supplemented by an angular component motivated by axial symmetry~\cite{Cunha:2020azh}. The vector field becomes singular at the poles, but its zero points—where the angular parameter is $\pi/2$ and $\tau = T^{-1}$—correspond to physically meaningful, on-shell black hole states.  Topologically, these zero points encode fixed points of the thermodynamic flow and allow one to assign a topological charge to each black hole configuration. Locally, each zero of the thermodynamic vector field constructed from the generalized free energy corresponds to an on-shell black hole solution, thereby assigning a well-defined winding number to each configuration. This winding number serves as a topological invariant characterizing the thermodynamic nature of the solution. \cite{Wei:2022dzw} reveals that a positive winding number indicates a thermodynamically stable black hole, whereas a negative value signals instability. Notably, the emergence or annihilation of such zero points—corresponding to bifurcation or coalescence of solutions—plays a crucial role in the dynamical evolution of black holes in a finite-temperature cavity, marking potential transitions between stable and unstable phases.
%%%%%%%%%%%%%%%%%%%%%%%%%%%%%%%%%%%%%%%%%%%%%%%%%%%%%%%%%%%%%%%%%%%%%%%%%%

\quad A notable advancement in black hole thermodynamics emerges after introducing the notion of ADM mass from asymptotically flat to anti-de Sitter $(AdS)$ spacetimes. With regard to this generalization, the cosmological constant $\Lambda$ must be treated as a thermodynamic variable\cite{Kubiznak:2012wp}. When reformulating Smarr's relation for AdS spacetimes, a dynamical $\Lambda$ is essential. This naturally leads to the interpretation of $\Lambda$ as a bulk pressure $P$ and the introduction of a conjugate thermodynamic volume $V$. This framework, wherein the first law of black hole thermodynamics includes a new $VdP$ term associated with a variable $\Lambda$, is referred to as extended black hole thermodynamics\cite{Kastor:2009wy,Cvetic:2010jb}. Within this framework, the ADM mass $M$ of the black hole is reinterpreted as its enthalpy $H$, rather than internal energy $U$, such that $M=H=U+PV$. Consequently, the differential form of the first law can be written as,
\begin{equation}\label{first law}
    dH = TdS + VdP,
\end{equation}
where the thermodynamic volume is defined as $V = (\partial H/ \partial P)_S$, conjugate to the thermodynamic pressure $P$. This formalism serves as a precise analogy between the phase transitions of charged black holes and the liquid-gas transitions of a van der Waals (vdW) fluid, thereby placing both systems within the same universality class.

\quad Knowing that a temperature can be attributed to black holes, this implies a potential microscopic description that is compatible with the Bekenstein-Hawking entropy, albeit with some degrees of freedom\cite{Bekenstein:1974ax,PhysRevD.13.191,Hawking:1975vcx}. An additional empirical method for examining phase transitions can be acquired by thermodynamic geometry, specifically through the application of Ruppeiner's thermodynamic curvature~\cite{RevModPhys.67.605}. This approach applies to almost any generic thermodynamic system in nature, alongside black holes. In any thermodynamic system, it serves as a useful diagnostic tool for establishing a general understanding of the nature of interactions among its microstructure. This method, which was first used by Weinhold \cite{Weinhold:1975xej} and Ruppeiner \cite{Ruppeiner:1979bcp,Ruppeiner:1995zz}, establishes a Riemannian metric on the manifold of extensive variables (such as energy, volume, and charge), allowing for the measurement of underlying interactions between microscopic constituents using the Ruppeiner curvature. A flat curvature indicates non-interacting behavior (such as an ideal gas), whereas divergences or changes in sign frequently indicate critical phenomena. This formalism has been effectively used for a variety of systems, such as magnetic materials, Ising models, and quantum liquids, and it has also been extended to gravitational contexts, such as AdS and Kerr black holes, where curvature singularities correspond to phase transitions or thermodynamic instability. Ruppeiner geometry is therefore ideally suited to investigate the possible effects of Lorentz invariance violation on microstructure interactions, which could result in curvature signatures that are different from relativistic cases. Further, these advancements have broadened the understanding of black hole thermodynamics in the AdS background.

The Event Horizon Telescope (EHT) collaboration’s images of the supermassive-black-hole shadows in M87* and Sagittarius A* mark a pivotal moment in black-hole physics for studying its optical features. Although an event horizon itself is invisible because it emits no light, the surrounding shadow—a dark silhouette against a bright background produced by extreme gravitational lensing—can be resolved observationally. Since those images appeared, efforts have multiplied on two complementary fronts: (i) improving interferometric imaging techniques for sharper shadows, and (ii) building theoretical templates of how shadows and particle orbits look in alternative theories of gravity.
In particular, Lorentz-violating extensions of general relativity introduce new couplings that reshape not only the photon sphere but also the timelike geodesic structure around black holes. Our Section \ref{Sec:Shadow} shows that the lorentz-violating parameters $\ell_{1,2}$ and the cosmological constant $\Lambda$ shift the innermost stable circular orbit (ISCO), create novel precessing and plunging trajectories, and modulate the range of bounded orbits available to massive particles. These effects can imprint themselves on stellar dynamics, hot-spot precession, and accretion-flow variability, furnishing observational tests that complement shadow measurements. Consequently, a complete picture now requires analysing both null and timelike paths. The shadow geometry still encodes the black hole’s mass, spin, charge, surrounding medium, and the observer’s line of sight. A non-rotating hole casts a circular shadow, whereas rotation distorts the boundary into a D-shaped figure whose asymmetry carries information about internal angular momentum and any deviations from Kerr geometry \cite{Vagnozzi:2022moj, Bambi:2019tjh, Chen:2022nbb, Johannsen:2010ru, Cunha:2017eoe, Vagnozzi:2020quf}. Yet the same Lorentz-violating couplings that modify the shadow also govern the motion of stars, gas clumps, and plasma blobs on timelike orbits. By jointly modelling these two observables, we can tighten constraints on Lorentz-violating gravity and sharpen predictions for forthcoming higher-resolution EHT campaigns.

The paper is organized as follows: In Section \ref{Sec:Black hole in Lorentz-Violating gravity}, we briefly review the magnetic black hole solution within the Lorentz-Violating gravity. In Subsection \ref{Subsec:Thermodynamics and Universality Relation}, we discuss the thermodynamics and universal relation of these black holes. In Section \ref{Sec:Thermodynamic Topology}, we discuss the thermodynamic topology and compute the topological charge. In Sec.~\ref {Sec:Photon Sphere}, we discuss the photon sphere. In Section \ref{Sec:Thermodynamic Topology}, we discuss the thermodynamic geometry and Ruppeiner geometry in these black holes. Sec.~\ref{Sec:Shadow} discusses the particle dynamics in the black-hole spacetime. Finally, section \ref{Sec:Conclusion} summarizes our results and findings.
%%%%%%%%%%%%%%%%%%%%%%%%%%%%%%%%%%%%%%%%%%%%%%%%%%%%%%%%%%
\section{Black hole in Lorentz-Violating gravity}\label{Sec:Black hole in Lorentz-Violating gravity}

We begin our analysis by considering the Einstein–Hilbert action~\cite{Altschul:2009ae} nonminimally coupled to a self-interacting Kalb–Ramond (KR) field
\begin{eqnarray}\label{S1}
S &=& \int d^{D}x \sqrt{-g} \left[ \frac{1}{16\pi}(R - 2\Lambda) - \frac{1}{12}H^{\mu\nu\rho} H_{\mu\nu\rho} - V(B^{\mu\nu} B_{\mu\nu} \pm b^{2}) + \frac{1}{16\pi} \left( \xi_{1} B^{\mu\nu} B_{\mu\nu} R + \xi_{2} B^{\rho\mu} B^{\nu}{}_{\mu} R_{\rho\nu} \right) \right] \ ,
\end{eqnarray}
where $\Lambda$ denotes the cosmological constant, and $\xi_1$, $\xi_2$ are the coupling constants governing the interaction strength between the gravitational field and the Kalb–Ramond field. The KR field $B_{\mu\nu}$ is a rank-2 antisymmetric tensor whose field strength is defined by
\begin{equation}
H_{\mu\nu\rho} \equiv \partial_{[\mu} B_{\nu\rho]} \ ,
\end{equation}
which is manifestly invariant under the gauge transformation
$B_{\nu\rho} \rightarrow B_{\nu\rho} + \partial_{\nu} \Lambda_{\rho} - \partial_{\rho} \Lambda_{\nu}$ with $\Lambda_\rho$ being an arbitrary 1-form field~\cite{Altschul:2009ae}. Following the gravitational sector of the Standard Model Extension ~\cite{Colladay:2001wk, Kostelecky:2000mm, Kostelecky:2001mb, Colladay:2009rb, Berger:2015yha, Carroll:1989vb, Andrianov:1994qv, Andrianov:1998wj, Lehnert:2004hq, BaetaScarpelli:2012kt, Brito:2013npa}, let's introduce a self-interaction potential for the KR field as
\begin{equation}
V = V(B^{\mu\nu} B_{\mu\nu} \pm b^2) \ ,
\end{equation}
where the nonvanishing vacuum expectation value (VEV) of the field is $\langle B_{\mu\nu} \rangle = b_{\mu\nu}$, constrained by the relation $b^{\mu\nu} b_{\mu\nu} = \mp b^2$. Now, we define $X = B^{\mu\nu} B_{\mu\nu} \pm b^2$ and $V'$ as the derivative w.r.t $x$ which will be utilized in subsequent computations. Following~\cite{Altschul:2009ae}, the KR tensor field may be decomposed as
\begin{equation}
B_{\mu\nu} = \tilde{E}_{[\mu} v_{\nu]} + \epsilon_{\mu\nu\alpha\beta} v^{\alpha} \tilde{B}^{\beta} \ ,
\end{equation}
where $v^{\mu}$ is a timelike vector, and $\tilde{E}_\mu$, $\tilde{B}_\mu$ are spacelike vectors interpreted respectively as pseudo-electric and pseudo-magnetic components and their contraction with $v^{\nu}$. Unlike the VEV of the bumblebee field, which yields a single background vector, the VEV of the KR field introduces two independent background vectors. For the sake of simplicity, we assume a configuration where only the pseudo-electric component is nonvanishing, i.e., $b_2 = -\tilde{E}(r) dt \wedge dr$ which in component notation becomes $b_{10} = -b_{01} = \tilde{E}(r)$.

Several studies have proposed that the nonminimal term $\xi_1 B^{\mu\nu} B_{\mu\nu} R$ can be effectively absorbed into the Einstein–Hilbert term via a redefinition, such as replacing it with $\mp \xi_1 b^2 R$ in vacuum~\cite{Yang:2023wtu, Lessa:2019bgi, Duan:2023gng}. However, a careful variational analysis reveals otherwise. Varying the term $B^{\mu\nu} B_{\mu\nu} R \sqrt{-g}$ with respect to the metric yields
\begin{eqnarray}\label{BBR}
\frac{\delta( B^{\alpha\beta}B_{\alpha\beta}R\sqrt{-g})}{\delta g^{\mu\nu}} &&= \sqrt{-g}\left[g_{\mu\nu} \nabla^{2}(B^{\alpha\beta} B_{\alpha\beta}) + B^{\alpha\beta} B_{\alpha\beta} G_{\mu\nu}  - 2 B^{\alpha}_{~\mu} B_{\nu\alpha} R - \nabla_{\mu} \nabla_{\nu}(B^{\alpha\beta} B_{\alpha\beta}) \right]  \ , \\
&&\frac{\delta(b^2 R \sqrt{-g})}{\delta g^{\mu\nu}} = \xi_1 b^2 G_{\mu\nu} \sqrt{-g} \ .
\label{bbR}
\end{eqnarray}
Clearly, even under the condition $B^{\mu\nu} B_{\mu\nu} = \mp b^2$, the expressions in Eqs.~\eqref{BBR} and \eqref{bbR} are not equivalent. This discrepancy highlights a critical oversight in earlier analyses that assumed a direct absorption of the nonminimal coupling term into the gravitational sector. In fact, the term $\xi_1 B^{\mu\nu} B_{\mu\nu} R$ contributes nontrivially to the gravitational field equations, and any attempt to eliminate it through field redefinitions fails to capture its full dynamical role.  Finally, Varying the action~\eqref{S1} w.r.t. metric yields the gravitational field equations as
\begin{eqnarray}
    &&G_{\mu\nu}+\Lambda g_{\mu\nu}= 8\pi \Bigg\{\frac{1}{2}H_{\mu\alpha\beta}H_{\nu}^{\alpha\beta}-\frac{1}{12}g_{\mu\nu}H^{\alpha\beta\rho}H_{\alpha\beta\rho}+\xi_{1}\big[\nabla_{\mu}\nabla_{\nu}(B^{\alpha\beta}B_{\alpha\beta})-g_{\mu\nu}\nabla^{2}(B^{\alpha\beta}B_{\alpha\beta})-B^{\alpha\beta}B_{\alpha\beta}G_{\mu\nu} \nonumber \\
		&& +2B^{\alpha}_{\mu}B_{\nu}{}_{\alpha}R\big]+\xi_{2}\biggl[\frac{1}{2}g_{\mu\nu}B^{\alpha\gamma}B^{\beta}{}_{\gamma}R_{\alpha\beta}-B^{\alpha}{}_{\mu}B^{\beta}{}_{\nu}R_{\alpha\beta}-B^{\alpha\beta}B_{\nu\beta}R_{\mu\alpha}-B^{\alpha\beta}B_{\mu\beta}R_{\nu\alpha} +\frac{1}{2}\nabla_{\alpha}\nabla_{\mu}\left(B^{\alpha\beta}B_{\nu\beta}\right) \nonumber \\
		&&+\frac{1}{2}\nabla_{\alpha}\nabla_{\nu}\left(B^{\alpha\beta}B_{\mu\beta}\right)-\frac{1}{2}\nabla^{\alpha}\nabla_{\alpha}\left(B_{\mu}^{\gamma}B_{\nu\gamma}\right)-\frac{1}{2}g_{\mu\nu}\nabla_{\alpha}\nabla_{\beta}\left(B^{\alpha\gamma}B^{\beta}{}_{\gamma}\right)\bigg]\Bigg\} +4V^{\prime}(X)B_{\alpha\mu}B^{\alpha}{}_{\nu}-g_{\mu\nu}V(X) \ .
\end{eqnarray}

 We consider a general metric ansatz describing a static, spherically symmetric, four-dimensional spacetime, given by
\begin{eqnarray} \label{ansatz}
    ds^{2} = -f(r)dt^{2} + \frac{dr^{2}}{f(r)} + r^{2} \left(d\theta^{2} + \sin^{2}\theta\, d\varphi^{2}\right) \ ,
\end{eqnarray},
The VEV ansatz for the Kalb–Ramond field is specified by Eq.~\eqref{ansatz}. Given the condition $g^{\mu\alpha}g^{\nu\beta}b_{\mu\nu}b_{\alpha\beta} = -b^2$, and employing the chosen metric ansatz, one can express the pseudo-electric component $\tilde{E}(r) =|b|/\sqrt{2} $. To simplify the notation and facilitate the analysis of Lorentz-violating effects, we define the effective coupling parameters
\begin{equation}
\ell_1 = b^2 \xi_1, \quad \ell_2 = b^2 \xi_2 \ ,
\end{equation}
which encapsulate the strength of nonminimal couplings between the Kalb–Ramond background and the spacetime curvature. The different choices of potential $V(x)$ give different black hole solutions~\cite {Liu:2025fxj}. For the linear potential function, the metric function is  \begin{eqnarray}\label{Metric Linear}
	f(r)&=&\frac{2(1+\ell_1)}{2+2\ell_1-\ell_2} -\frac{2 M}{r}-\frac{2\Lambda }{3(2-2\ell_1-\ell_2)}r^2 \ .
\end{eqnarray}
For the Quadratic choice, the metric function is 
\begin{eqnarray}\label{Metric Quadratic}
	f(r)&=&\frac{(1+\ell _1)}{1+3\ell _1} -\frac{2 M}{r}-\frac{\Lambda }{3(1+\ell_1)}r^2 \ .
\end{eqnarray}
In particular, Lorentz-violating extensions of general relativity introduce new couplings that reshape not only the photon sphere but also the timelike geodesic structure around black holes.

%---------------------------------
\subsection{Thermodynamics and Universality Relation}\label{Subsec:Thermodynamics and Universality Relation}
In this subsection, we discuss thermodynamics and universal relations for these black holes. To examine the validity of the universality relation, we introduce a perturbative modification to the action, proportional to the cosmological constant. This perturbation induces corrections not only in the spacetime geometry but also in the associated thermodynamic quantities. By incorporating a perturbative term—parameterized by $\varepsilon$ and scaled with respect to the cosmological constant—we can systematically compute the corrected mass and other relevant thermodynamic variables. Utilizing these perturbed expressions, the universality relation given in Eq.~\eqref{Extremality condition} can be readily tested and verified.

%-------------------------
\subsubsection{Linear Case}

From Eq.~\eqref{Metric Linear} and perturbing the action~\eqref{S1} with perturbation parameter $\varepsilon$, the perturbed mass is 
\begin{eqnarray}\label{M linear}
    M(\varepsilon) =\frac{\sqrt{S}}{3 \pi ^{3/2}} \left(\frac{3 \pi  (\ell_1+1)}{2\ell_1-\ell_2+2}-\frac{S(\epsilon +1)}{L^2 (2\ell_1+\ell_2-2){}}\right) \ .
\end{eqnarray}
Now, equating this with mass of the black hole the perturbation parameter $\varepsilon$ is 
\begin{eqnarray}\label{Epsilon Linear}
    \varepsilon =  \frac{3 \pi  L^2 (2\ell_1+\ell_2-2) \left(\sqrt{\pi } M (-2\ell_1+\ell_2-2)+(\ell_1+1) \sqrt{S}\right)}{S^{3/2} (2\ell_1-\ell_2+2)}-1 \ .
\end{eqnarray}
The perturbation parameter is 
\begin{eqnarray}\label{T linear}
    T(\varepsilon)= \frac{1}{2 \pi ^{3/2} \sqrt{S}}\left[\frac{\pi  (\ell_1+1)}{2\ell_1-\ell_2+2}-\frac{S (\epsilon +1)}{L^2 (2\ell_1+\ell_2-2)}\right] \ .
\end{eqnarray}
Using Eq.~\eqref{Epsilon Linear} and Eq.~\eqref{T linear}, it is easuy to verify the R.H.S. of the Eq.~\eqref{Extremality condition} we have 
\begin{eqnarray}\label{Tds linear}
    \lim_{M \to M_{\text{ext}}} -T \left( \frac{\partial S(M, \vec{\mathcal{Q}}, \varepsilon)}{\partial \varepsilon} \right)_{M, \vec{\mathcal{Q}}} = \frac{S^{3/2}}{3 \pi ^{3/2} L^2 (2\ell_1+\ell_2-2)} \ .
\end{eqnarray}
Finally Expanding Eq.~\eqref{M linear} in the powers of $\varepsilon$ we have 
\begin{eqnarray}
    M(\varepsilon) = M-\frac{S^{3/2} \epsilon }{3 \left(\pi ^{3/2} L^2 (2\ell_1+\ell_2-2)\right)} \ .
\end{eqnarray}
It is easy to check using this and Eq.~\eqref{Tds linear}, the extremality condition is verified in the Lorentz-Violating gravity.
%-------------------------
\subsubsection{Quadratic Case}

By introducing a perturbative deformation in the action \eqref{S1}, governed by the small parameter $\varepsilon$, and using the metric expansion given in Eq.~\eqref{Metric Quadratic}, the corresponding correction to the black hole mass is obtained as
\begin{eqnarray}\label{M Quadratic}
M(\varepsilon) = \frac{\sqrt{S} \left(\pi  L^2 (\ell_1+1)^2+(3 \ell_1+1) S (\epsilon +1)\right)}{2 \pi ^{3/2} L^2 (\ell_1+1) (3 \ell_1+1)} \ .
\end{eqnarray}

Solving for $\varepsilon$ by equating the perturbed mass to the physical black hole mass $M$, we obtain
\begin{eqnarray}\label{Epsilon Quadratic}
\varepsilon = \frac{2 \pi ^{3/2} L^2 (\ell_1+1) M}{S^{3/2}}-\frac{\pi  L^2 (\ell_1+1)^2}{3 \ell_1 S+S}-1 \ .
\end{eqnarray}

The temperature under this perturbation is similarly modified and reads
\begin{eqnarray}\label{T Quadratic}
T(\varepsilon) = \frac{\pi  L^2 (\ell_1+1)^2+3 (3 \ell_1+1) S (\epsilon +1)}{4 \pi ^{3/2} L^2 (\ell_1+1) (3 \ell_1+1) \sqrt{S}} \ .
\end{eqnarray}

Substituting the expressions for $\varepsilon$ and $T(\varepsilon)$ into the right-hand side of the extremality identity \eqref{Extremality condition}, we find
\begin{eqnarray}\label{Tds Quadratic}
\lim_{M \to M_{\text{ext}}} -T \left( \frac{\partial S(M, \vec{\mathcal{Q}}, \varepsilon)}{\partial \varepsilon} \right)_{M, \vec{\mathcal{Q}}} = -\frac{S^{3/2}}{2 \pi ^{3/2} L^2 (\ell_1+1)} \ .
\end{eqnarray}
Expanding the perturbed mass in a Taylor series around $\varepsilon = 0$, we obtain the leading-order correction as
\begin{eqnarray}
M(\varepsilon) = M +\frac{S^{3/2} \epsilon }{2 \pi ^{3/2} L^2 (\ell_1+1)} \ .
\end{eqnarray}
This linear variation of the mass with respect to the perturbation parameter $\varepsilon$ is consistent with the result from Eq.~\eqref{Tds Quadratic}, thereby confirming that the universal extremality relation holds true in the framework of Lorentz-Violating gravity-modified gravity with Lorentz-violating corrections.

%%%%%%%%%%%%%%%%%%%%%%%%%%%%%%%%%%%%%%%%%%%%%5
\section{Thermodynamic Topology}\label{Sec:Thermodynamic Topology}

The exploration of black hole thermodynamics has increasingly embraced topological methodologies to elucidate complex phase structures that elude conventional analysis via standard thermodynamic potentials. A particularly influential contribution in this context is Duan’s $\varphi$-mapping topological current theory \cite{Duan:1979ucg, Duan:1984ucg}, which provides a robust mathematical framework for identifying critical phenomena. Within this formalism, phase transitions are interpreted through the emergence of topological defects, defined as the zeros of a vector field $\phi$. These singularities give rise to a conserved topological current, mathematically derived from a generalized Jacobian tensor. Under regularity conditions, this construction simplifies, with the topological current expressible as the determinant of the Jacobian matrix associated with the $\phi$-mapping. This topological perspective offers deep insights into the nature and classification of black hole phase transitions. The associated topological invariant, typically denoted by $W$, encapsulates the global features of the thermodynamic phase landscape and is evaluated through a decomposition involving the Hopf index and Brouwer degree. These topological indices measure the algebraic winding number of the vector field $\boldsymbol{\phi}$ around its isolated zeros, thereby encoding crucial information about the underlying defect structure \cite{Wei:2022dzw, Wei:2021vdx}. A positive value of the topological charge corresponds to a thermodynamically stable black hole configuration, whereas a negative charge indicates local instability, often associated with bifurcations or metastable phases within the extended phase space. This formalism not only broadens the scope of conventional thermodynamic analysis but also forges a deep link between gravitational thermodynamics and topological field theory, offering a geometric framework for classifying black hole solutions in diverse spacetime geometries.

\begin{figure}[h!]
 \begin{center}
 \subfigure[]{
 \includegraphics[height=4.3cm,width=6cm]{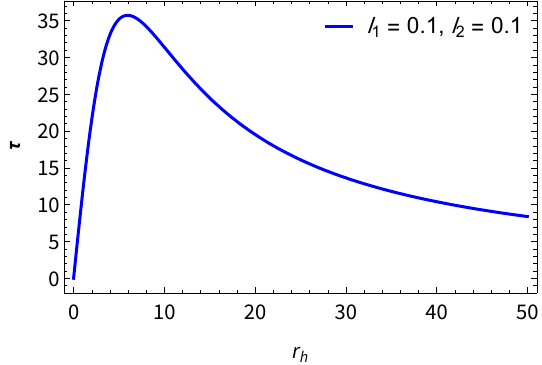}
 \label{3a}}
 \subfigure[]{
 \includegraphics[height=4.3cm,width=6cm]{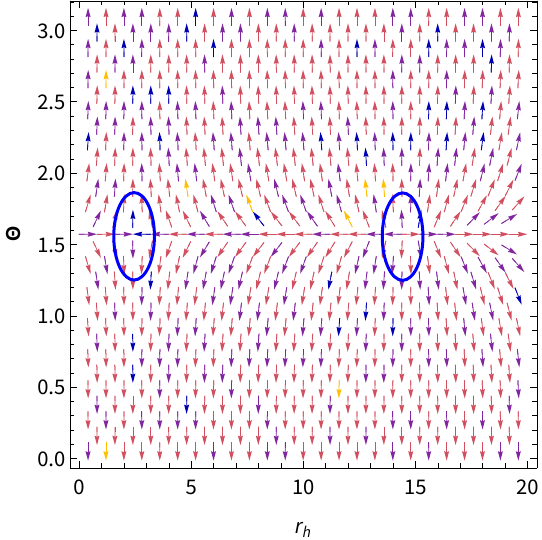}
 \label{3b}}
 \subfigure[]{
 \includegraphics[height=4.3cm,width=6cm]{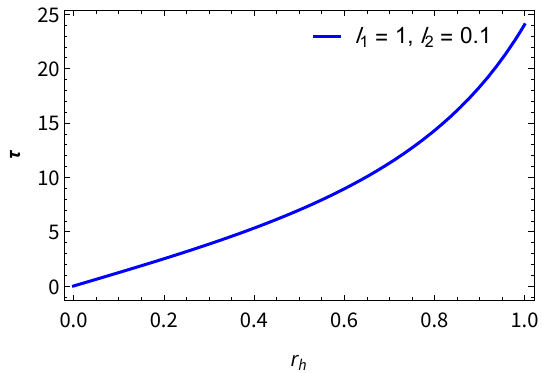}
 \label{3c}}
 \subfigure[]{
 \includegraphics[height=4.3cm,width=6cm]{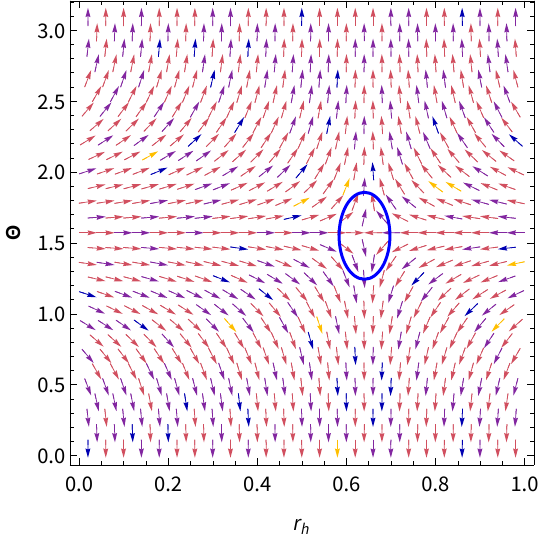}
 \label{3d}}
 \subfigure[]{
 \includegraphics[height=4.3cm,width=6cm]{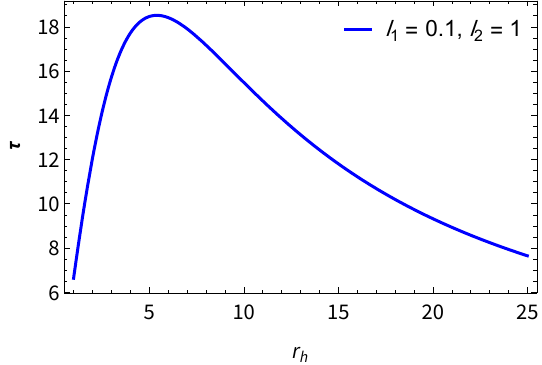}
 \label{3e}}
 \subfigure[]{
 \includegraphics[height=4.3cm,width=6cm]{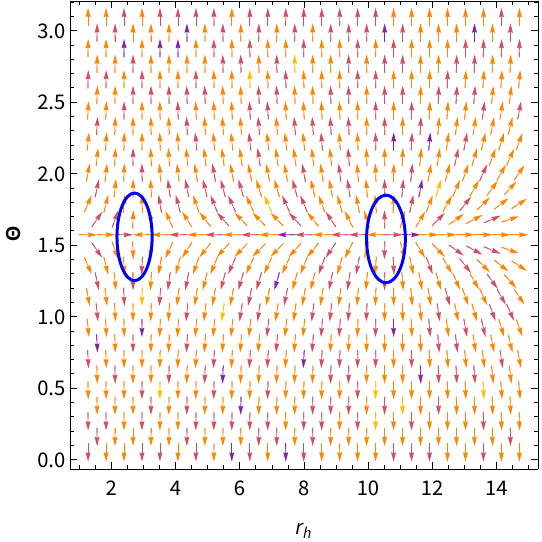}
 \label{3f}}
 \subfigure[]{
 \includegraphics[height=4.3cm,width=6cm]{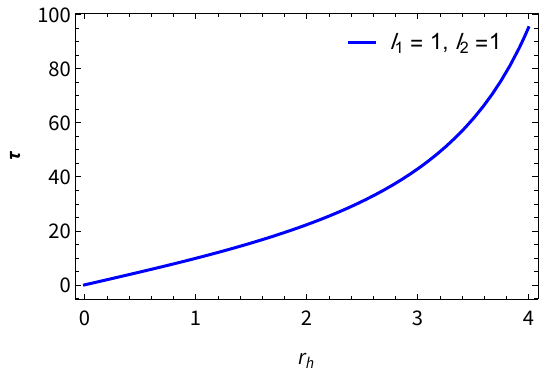}
 \label{3g}}
 \subfigure[]{
 \includegraphics[height=4.3cm,width=6cm]{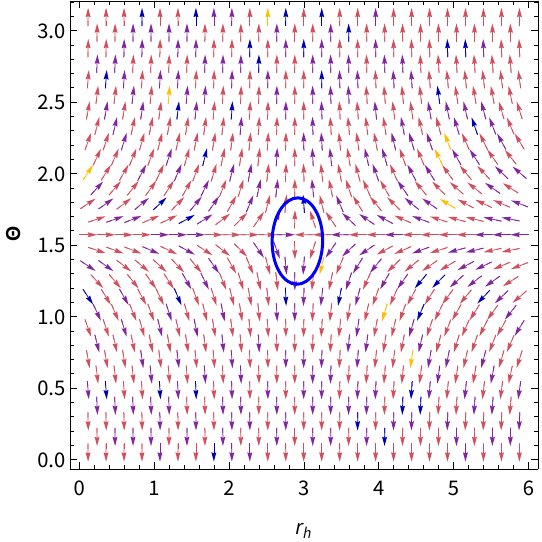}
 \label{3h}}
 \caption{\small{The \((\tau \text{ vs. } r_h)\) diagram, depicted in Figs. (\ref{3a}), (\ref{3c}), (\ref{3e}), and (\ref{3g}), illustrates the variations in free parameters for a black hole with a nonlinear electromagnetic field in the presence of a phantom global monopole. This representation elucidates the dependence of thermodynamic quantities on the horizon radius \( r_h \), highlighting critical transition points within the phase structure. Additionally, the normal vector field \( n \) in the \((r_h - \Theta)\) plane is presented, demonstrating the distribution of Zero Points (ZPs) at specific coordinates \((r_h, \Theta)\). These ZPs correspond to parameter values \(\ell_1 = 0.1, 1\) and \(\ell_2 = 0.1, 1\), serving as key indicators of stability properties and topological characteristics within the thermodynamic framework.}}
 \label{m3}
\end{center}
 \end{figure}
 
The topological approach to black hole thermodynamics equips the phase space with a generalized free energy structure, derived from the Euclidean path integral formalism. In this framework, equilibrium states are identified by enforcing the periodicity condition in the Euclidean time coordinate $\tau$, which must be the inverse of the Hawking temperature $T^{-1}$. This requirement ensures that the configuration remains on-shell and physically consistent within semiclassical gravity \cite{Wei:2022dzw, Wei:2021vdx, a19,a20,20a,21a,22a,23a,24a,26a,27a, 28a, 29a, 31a, 33a, 34a, 35a, 37a, 38', 38a, 38b, 38c, 39a,40a,42a, 43a, 44a, 44c, 44d, 44e, 44f, 44g, 44h, 44i, 44j, 44k, 44l, 44m, 44n, 44o, 44p}. The generalized Helmholtz free energy is given by
\begin{eqnarray}\label{Helmholtz Energy}
    \mathscr{F} = M - \frac{S}{\tau} \ , 
\end{eqnarray}
where $M$ is the ADM mass of the black hole, $S$ its entropy, and $\tau$ the Euclidean time period. The free energy reduces to its on-shell form when the temperature is identified as $T = \tau^{-1}$. To analyze the critical behavior embedded in this free energy landscape, a two-component vector field $\boldsymbol{\phi}$ is introduced, capturing variations with respect to thermodynamic parameters. A representative form is
\begin{eqnarray}\label{Phi Definition}
    \boldsymbol{\phi} = \left( \frac{\partial \mathscr{F}}{\partial r_h},\ -\cot \Theta \csc \Theta \right) \ ,
\end{eqnarray}
where $r_h$ denotes the horizon radius and $\Theta \in [0, \pi]$ is an auxiliary angle introduced to parametrize the manifold over which the vector field is defined. Notably, the second component diverges at $\Theta = 0$ and $\Theta = \pi$, pushing the vector field outward at these boundary points. Within the formalism of Duan’s $\varphi$-mapping theory, one constructs a conserved topological current supported only at the zeros of the vector field. This current is defined as
\begin{eqnarray}
    j^\mu = \frac{1}{2\pi} \epsilon^{\mu\nu\rho} \epsilon^{ab} \partial_\nu n^a \partial_\rho n^b, \quad n^i = \frac{\phi^i}{\|\phi\|} \ , 
\end{eqnarray}
with $\mu, \nu, \rho = 0,1,2$ and $i = r_h, \Theta$. The current $j^\mu$ vanishes everywhere except at isolated points where $\boldsymbol{\phi} = 0$, indicating thermodynamic criticality. The total topological charge~\cite{Wei:2020rbh}, which serves as a global invariant of the phase structure, is given by
\begin{eqnarray}
    W = \int_\Sigma j^0 \, d^2x = \sum_{i=1}^{n} \zeta_i \eta_i = \sum_{i=1}^{n} \omega_i \ ,
\end{eqnarray}
where $\zeta_i$ is the Hopf index counting the multiplicity of the zero at point $z_i$, and $\eta_i = \pm 1$ is the Brouwer degree indicating the local orientation of the mapping. The resulting winding number $\omega_i = \zeta_i \eta_i$ provides a quantitative classification of each topological defect. This framework has been applied successfully to a broad class of gravitational systems, including asymptotically AdS/dS black holes, and models incorporating non-linear electrodynamics, higher-curvature terms, or exotic scalar fields. In the subsequent analysis, we apply this method to study black holes in the Lorentz-violating gravity case. The generalized Helmholtz free energy obtained from Eq.~\eqref{Helmholtz Energy} forms the basis for topologically probing their phase structure.

%------------------------------------------
\subsection{Linear Case}

The generalized Helmholtz free energy for the first case can be written as 
\begin{eqnarray}\label{Vector Field First}
    \mathscr{F} = -\frac{8 \pi  P r_h^3}{3 (2 \ell_1+\ell_2-2)}+\frac{(\ell_1+1) r_h}{2 \ell_1-\ell_2+2}-\frac{\pi  r_h^2}{\tau }.
\end{eqnarray}
Using Eq.~\eqref{Phi Definition}, we can compute the vector field $\phi$ as 
\begin{eqnarray}
    \phi^{r_h} = -\frac{8 \pi  P r_h^2}{2 \ell_1+\ell_2-2}+\frac{\ell_1+1}{2 \ell_1-\ell_2+2}-\frac{2 \pi  r_h}{\tau }\;\;\;\text{and}\;\;\; \phi ^{\theta } = -\cot{\theta}\csc{\theta} \,
\end{eqnarray}
Upon solving Eq.~\eqref{Vector Field First}, we obtain the following expression
\begin{eqnarray}\label{Tau first case}
    \tau = \frac{2 \pi  \left(4 \ell_1^2 r_h-\ell_2^2 r_h+4 \ell_2 r_h-4 r_h\right)}{2 \ell_1^2+\ell_1 \ell_2-16 \pi  \ell_1 P r_h^2+8 \pi  \ell_2 P r_h^2+\ell_2-16 \pi  P r_h^2-2};
\end{eqnarray}

%------------------------------------------
\begin{figure}[h!]
 \begin{center}
 \subfigure[]{
 \includegraphics[height=4.3cm,width=6cm]{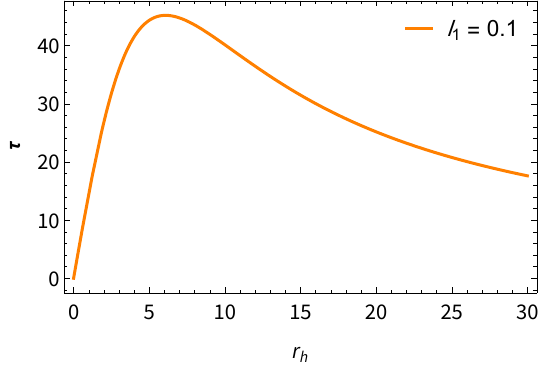}
 \label{4a}}
 \subfigure[]{
 \includegraphics[height=4.3cm,width=6cm]{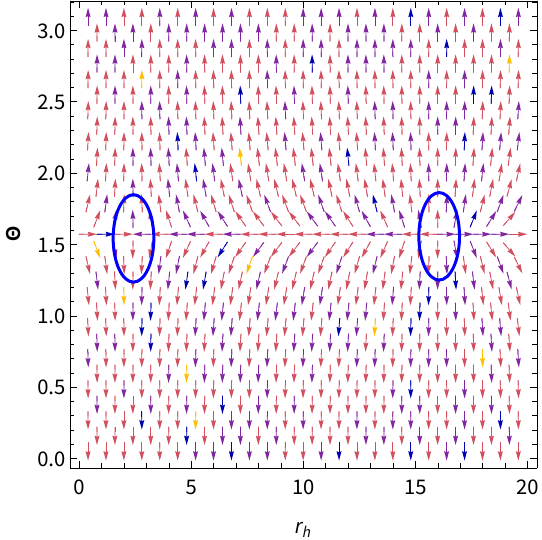}
 \label{4b}}
 \subfigure[]{
 \includegraphics[height=4.3cm,width=6cm]{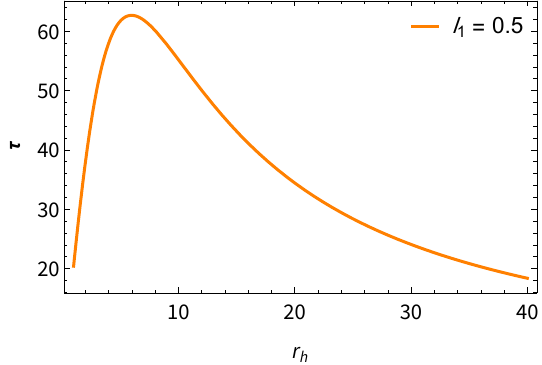}
 \label{4c}}
 \subfigure[]{
 \includegraphics[height=4.3cm,width=6cm]{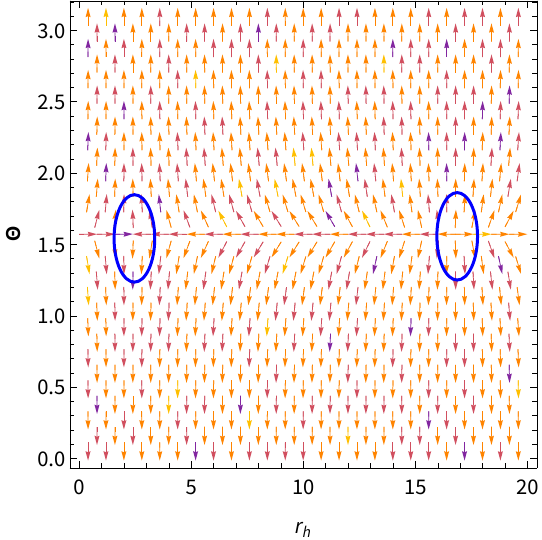}
 \label{4d}}
 \subfigure[]{
 \includegraphics[height=4.3cm,width=6cm]{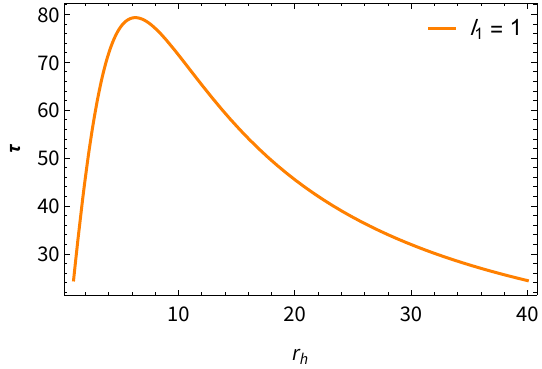}
 \label{4e}}
 \subfigure[]{
 \includegraphics[height=4.3cm,width=6cm]{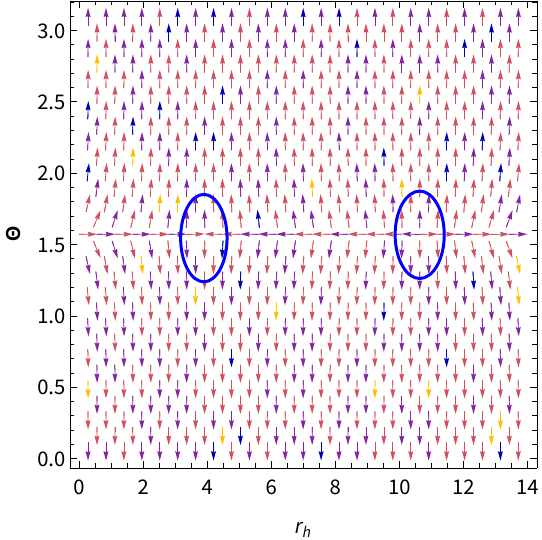}
 \label{4f}}
 \caption{\small{The \((\tau \text{ vs. } r_h)\) diagram, as illustrated in Figs. (\ref{4a}), (\ref{4c}), and (\ref{4e}), depicts the variation of free parameters associated with a black hole characterized by a nonlinear electromagnetic field in the presence of a phantom global monopole. This graphical representation provides insight into the dependence of thermodynamic quantities on the horizon radius \( r_h \), highlighting crucial transition points within the phase structure. Furthermore, the normal vector field \( n \) in the \((r_h - \Theta)\) plane is introduced, illustrating the spatial distribution of Zero Points (ZPs) at specific coordinates \((r_h, \Theta)\). These ZPs, corresponding to parameter values \(\ell_1 = 0.1, 0.5, 1\), play a fundamental role in revealing the stability properties and topological characteristics of the thermodynamic system. }}
 \label{m4}
\end{center}
 \end{figure}

\subsection{Quadratic Case}

The generalized Helmholtz free energy for the second case can be written as 
    \begin{eqnarray}
    \mathscr{F} = \frac{4 \pi  P r_h^3}{3 (\ell_1+1)}+\frac{(\ell_1+1) r_h}{6 \ell_1+2}-\frac{\pi  r_h^2}{\tau }\ .
\end{eqnarray}
Utilizing Eq.~\eqref{Phi Definition}, the components of the vector field $\boldsymbol{\phi}$ are computed as:
\begin{eqnarray}\label{Vector Field Second}
\phi^{r_h} = \frac{4 \pi  P r_h^2}{\ell_1+1}+\frac{\ell_1+1}{6 \ell_1+2}-\frac{2 \pi  r_h}{\tau }\;\;\;\text{and}\;\;\;
\phi^{\theta} = -\cot{\theta} \csc{\theta} \ .
\end{eqnarray}
Substituting these components into Eq.~\eqref{Vector Field Second} and solving accordingly, we arrive at the following analytical expression for $\tau$
\begin{eqnarray}\label{Tau Second case}
\tau = \frac{4 \pi  \left(3 \ell_1^2 r_h+4 \ell_1 r_h+r_h\right)}{\ell_1^2+24 \pi  \ell_1 P r_h^2+2 \ell_1+8 \pi  P r_h^2+1} \ .
\end{eqnarray}

We conduct a comprehensive investigation into the thermodynamic topology of black holes in Lorentz-Violating gravity, focusing on the distribution of topological charges as depicted in Fig.~\ref {m3} (linear case) and Fig.~\ref {m4} (quadratic case). The normalized field lines illustrated in these figures provide a detailed visualization of the underlying topological structure. Specifically, Fig.~(\ref{m3}) identifies two distinct topological classifications: one with a single zero point and another with two distinct zero points at specific coordinates \((r_h, \Theta)\). These zero points, corresponding to parameter values \( \ell_1 = 0.1, 1 \) and \( \ell_2 = 0.1, 1 \), represent localized topological charges enclosed within blue contour loops, whose configurations are governed by variations in the free parameters.
A key result from this analysis is that, independent of parameter modifications, the system maintains two topological charges, \( (\omega = +1, -1) \). This classification remains unchanged across different parameter selections, yielding a total topological charge of \( W = 0 \), as illustrated in Fig.~\ref{3b} and Fig.~\ref{3f}. Conversely, in cases where the system exhibits a single topological charge, \( (\omega = -1) \), the total topological charge is found to be \( W = -1 \), as demonstrated in Fig.~\ref{3d} and Fig.~\ref{3h}. The stability of the black hole is further examined through winding number calculations, reinforcing this classification. However, under specific parameter variations, distinct topological charge distributions emerge. As shown in Fig.~(\ref{m4}) (Case II), when the parameter set \(\ell_1 = 0.1, 0.5, 1\) is considered, the system again exhibits two topological charges, \( (\omega = +1, -1) \), resulting in a total charge of \( W = 0 \), as represented in Fig.~(\ref{m4}). 
To provide further insight into this topological framework, we analyze the free energy function as a scalar quantity mapped within the two-dimensional space \((r_h, \Theta)\). The corresponding vector field \( \phi \) is structured so that the extremum points of the free energy function align with the zero points of this field. The rotational behavior of field lines surrounding these zero points—determined by their association with either maxima or minima—offers a systematic approach for assigning topological charges \cite{a19}.
A comparative study of fundamental black hole solutions, including Schwarzschild and Reissner-Nordström configurations, reveals distinct trends in their respective topological charge values: Schwarzschild black hole: \( W = -1 \); Reissner-Nordström black hole: \( W = 0 \); AdS-Reissner-Nordström black hole: \( W = +1 \) \cite{a19}. These classical solutions serve as fundamental models in black hole thermodynamics, facilitating a broader classification of black hole structures and their thermodynamic properties. In the context of our study, the results are consistent with established findings, as verified in \cite{a19}. This topological approach not only validates theoretical predictions but also provides a structured framework for assessing black hole stability and phase transitions, further extending its applicability to gravitational thermodynamics and astrophysical phenomena.
%%%%%%%%%%%%%%%%%%%%%%%%%%%%%%%%%%%%%%%%%%%%%%%%%%%%%%%%%%%%%%%%%%%%
%%%%%%%%%%%%%%%%%%%%%%%%%%%%%%%%%%%%%%%%%%%%%%%%%%%%%%%%%%

\section{Photon Sphere}\label{Sec:Photon Sphere}

Photon spheres represent critical hypersurfaces composed of closed null geodesics, delineating the region where gravitational lensing becomes non-perturbative. They form the innermost boundary for photon orbits and manifest as either unstable configurations—essential to the observable black hole shadow—or as stable structures often linked to dynamical instabilities in spacetime. Conventionally, these structures are identified via extrema of an effective potential derived from conserved quantities of geodesic motion. However, this methodology is inherently tied to particle-specific parameters.

To overcome such limitations, we employ a covariant, geometry-driven formulation that encapsulates the local behavior of photon trajectories independent of particle characteristics. This involves defining a scalar potential $\mathcal{H}(r,\theta)$, constructed solely from the spacetime metric, specifically tailored for null geodesics
\begin{eqnarray}\label{H function}
    \mathcal{H}(r,\theta) = \frac{1}{\sin\theta} \left( \frac{f(r)}{r^2} \right)^{1/2} \,.
\end{eqnarray}
This function encodes the intrinsic spacetime geometry and facilitates a reduced-dimensional analysis on the equatorial plane, mapping the system into a 2D configuration analogous to a Poincaré section. We define a vector field $\phi^a = (\phi^r, \phi^\theta)$ on this spatial slice, constructed from the covariant derivatives of $\mathcal{H}(r,\theta)$ normalized by the metric components
\begin{eqnarray}\label{Vector Field Def.}
    \phi^r = \sqrt{f(r)}\,\partial_r \mathcal{H}, \quad \phi^\theta = \frac{1}{r}\,\partial_\theta \mathcal{H} \ .
\end{eqnarray}
These components are compactly represented in polar form as $\phi = |\phi|e^{i\Theta}$, with the norm $|\phi| = \sqrt{\phi^a \phi\_a}$ and orientation $\Theta$ governing the local directionality of the flow. The associated unit vector field $n^a = \phi^a/|\phi|$ describes the normalized flow of photon trajectories in this configuration space.

Topological features of this flow are encoded via the topological current
\begin{eqnarray}
    j^\mu = J^\mu(X)\, \delta^2(\phi), \quad Q = \int_\Omega J^0(X)\, \delta^2(\phi)\, d^2x \ ,
\end{eqnarray}
where $X$ denotes the Jacobian determinant of the mapping $\phi: \mathbb{R}^2 \rightarrow \mathbb{R}^2$, and $\delta^2(\phi)$ localizes the current at the critical points $\phi = 0$. These points correspond to photon spheres and are identified as topological defects with non-trivial index—analogous to vortices or monopoles in field theory. Importantly, the photon sphere's existence and multiplicity are now governed by the topology of $\phi^a$, rather than the dynamical characteristics of individual photons. This framework provides a unified, particle-independent approach to classify photon trapping surfaces across a wide range of black hole geometries, including those with AdS asymptotics or non-trivial matter couplings.

%---------------------------------------------------
\subsection{Linear Case}
The $\mathcal{H}-$function~\eqref{H function} for the linear case is 
\begin{eqnarray}\label{H-Function Linear}
    \mathcal{H}(r,\theta) = \frac{\csc{\theta}}{r} \sqrt{\frac{r^2}{3 L^2 \left(-\ell_1-\frac{\ell_2}{2}+1\right)}+\frac{\ell_1+1}{\ell_1-\frac{\ell_2}{2}+1}-\frac{2 M}{r}} \ .
\end{eqnarray}
Using Eq.~\eqref{Vector Field Def.} and Eq.~\eqref{H-Function Linear} the vector fields are 
\begin{eqnarray}
    \phi^{r_h} = \frac{\csc (\theta ) (M (6 \ell_1-3 \ell_2+6)-2 (\ell_1+1) r_h)}{r_h^3 (2 \ell_1-\ell_2+2)} \;\;\;\;;\;\;\;\; \phi^\theta = -\frac{\cot{\theta} \csc{\theta} }{r_h^2}\sqrt{-\frac{2 r_h^2}{3 L^2 (2 \ell_1+\ell_2-2)}+\frac{\ell_1+1}{\ell_1-\frac{\ell_2}{2}+1}-\frac{2 M}{r_h}}
\end{eqnarray}
\begin{figure}[h!]
 \begin{center}
 \subfigure[]{
 \includegraphics[height=6.3cm,width=8.5cm]{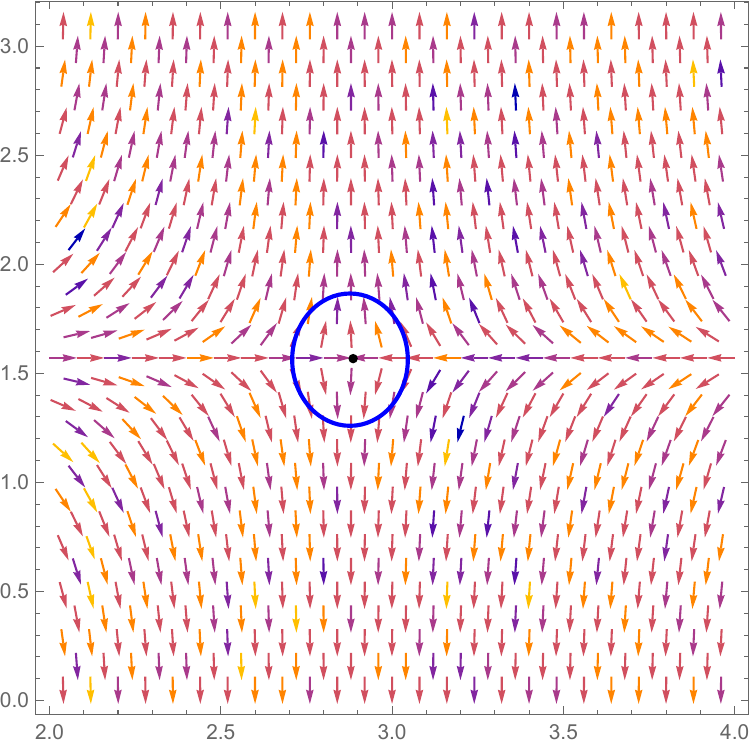}
 \label{PS1a}}
 \subfigure[]{
 \includegraphics[height=6.3cm,width=8.5cm]{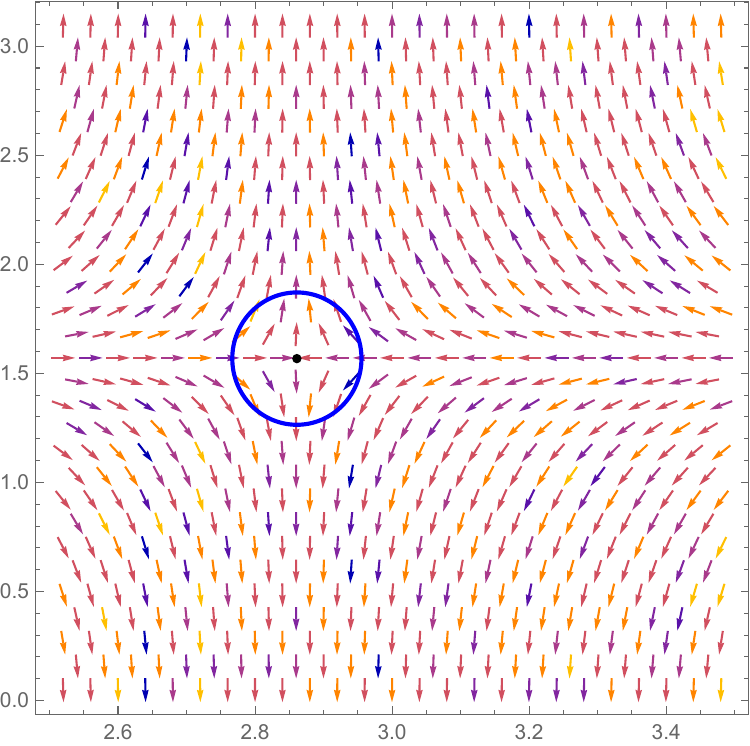}
 \label{PS1b}}
 \caption{\small{The photon spheres (PSs) for different parameter configurations with $\ell_1 = 0.1,  \ell_2=0.1$, \( M = 0.1, 1 \).}}
 \label{PS first case}
\end{center}
 \end{figure}

%---------------------------------------------------
\subsection{Quadratic Case}

The $\mathcal{H}-$function~\eqref{H function} for the quadratic case is 
\begin{eqnarray}\label{H-Function Quadratic}
    \mathcal{H}(r,\theta) = \frac{\csc (\theta ) \sqrt{\frac{r^2}{L^2 (\ell_1+1)}+\frac{\ell_1+1}{3 \ell_1+1}-\frac{2 M}{r}}}{r} \ .
\end{eqnarray}
Using Eq.~\eqref{Vector Field Def.} and Eq.~\eqref{H-Function Quadratic} the vector fields are 
\begin{eqnarray}
    \phi^{r_h} = -\frac{\csc (\theta ) (-9 \ell_1 M+\ell_1 r_h-3 M+r_h)}{(3 \ell_1+1) r_h^3} \;\;\;\;;\;\;\;\; \phi^\theta = -\frac{\cot (\theta ) \csc (\theta ) \sqrt{\frac{r_h^2}{L^2 (\ell_1+1)}+\frac{\ell_1+1}{3 \ell_1+1}-\frac{2 M}{r_h}}}{r_h^2}
\end{eqnarray}
\begin{figure}[h!]
 \begin{center}
 \subfigure[]{
 \includegraphics[height=6.3cm,width=8.5cm]{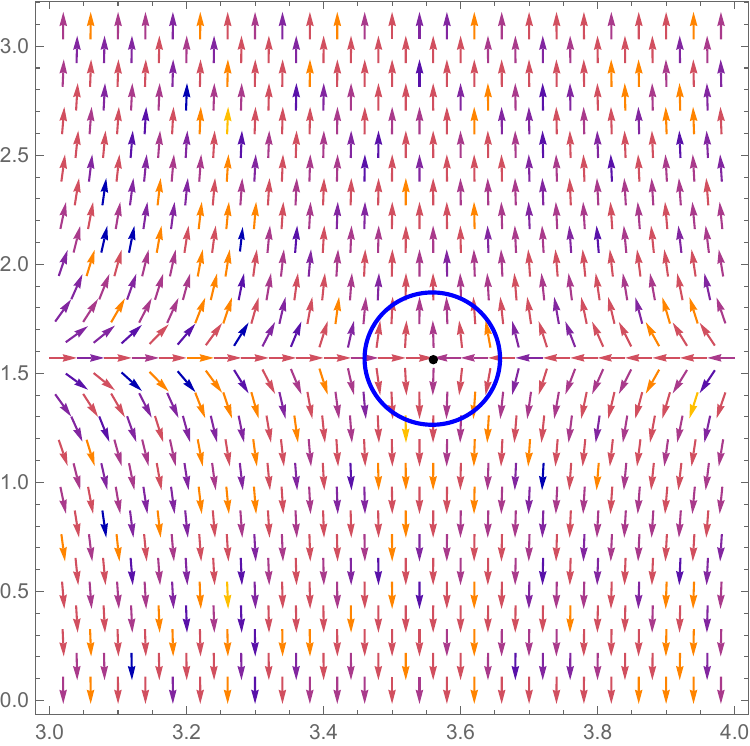}
 \label{PS2a}}
 \subfigure[]{
 \includegraphics[height=6.3cm,width=8.5cm]{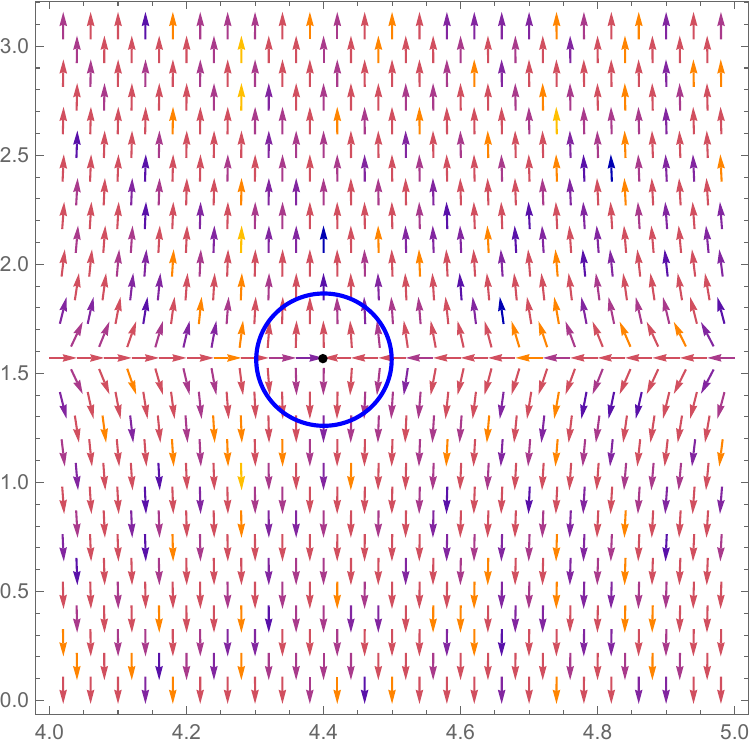}
 \label{PS2b}}
 \caption{\small{The photon spheres (PSs) for different parameter configurations with $\ell_1 = 0.1, 0.5$, \( M = 1 \).}}
 \label{PS Second case}
\end{center}
 \end{figure}

Through this geometric and topological framework, photon spheres are not only located with high precision but also characterized in terms of their stability and global significance within the spacetime. This approach has proven effective across various gravitational backgrounds and offers a deeper insight into null geodesic structures near ultra-compact objects. Within the framework of thermodynamic topology, photon spheres emerge as critical topological defects associated with the spacetime geometry near compact objects. Each zero of the vector field $\vec{\phi}$—corresponding to a photon sphere configuration—carries a topological charge, determined by the winding number of the field around the zero. This charge is quantized and can assume values of $+1$ or $-1$, contingent upon the local orientation and winding structure of the field. The total topological charge enclosed by a closed curve surrounding one or more such zero points is constrained to discrete values: $-1$, $0$, or $+1$. In charged static, spherically symmetric black hole solutions with $M > Q$, this topological analysis consistently yields a net photon sphere charge of $\text{PS} = -1$, reflecting the inherent instability of the photon orbit. Such topological characterization provides a powerful diagnostic for understanding the nature and stability of photon spheres across a broad class of gravitational systems.

To evaluate how this topology evolves with varying black hole parameters, we conduct an investigation of photon sphere behavior across multiple configurations. Specifically, we analyze cases for \( M = 0.1, 1 \), along with parameter variations \( \ell_1 = 0.1, 0.5 \) and \( \ell_2 = 0.1 \). The results, illustrated in Fig.~(\ref{PS first case}) and Fig.~(\ref{PS Second case}), demonstrate that for all examined configurations, the photon sphere charge consistently remains \( W = -1 \), reaffirming the robustness of its topological properties against perturbations in system parameters.

Further exploration into photon sphere structures provides crucial insights into the geometric and thermodynamic features of black hole spacetimes. By analyzing the orbital dynamics of trapped photons, we investigate how changes in coupling parameters \( \ell_1 \) and \( \ell_2 \) influence the structural stability of the photon sphere. This assessment extends to determining the critical points where photon trajectories reveal shifts in stability regimes. The interplay between photon sphere topology and black hole thermodynamic charges presents intriguing applications in gravitational optics, particularly in lensing phenomena and shadow formation. Through a systematic evaluation of entropy parameter variations, we construct a comprehensive framework to classify stable and unstable black hole configurations. The results consistently suggest that positive winding numbers are associated with thermodynamically stable solutions, while negative winding numbers indicate instability. This topological methodology not only refines existing theoretical models but also contributes to broader studies of black hole stability and phase transitions across diverse gravitational backgrounds.
%%%%%%%%%%%%%%%%%%%%%%%%%%%%%%%%%%%%%%%%%%%%%%%%%%%%%%%%%%%%%%%%%%%%%%%%%%%%

%%%%%%%%%%%%%%%%%%%%%%%%%%%%%%%%%%%%%%%%%%%%%%%%%%%%%%%%%

%%%%%%%%%%%%%%%%%%%%%%%%%%%%%%%%%%%%%%%%%%%%%%%%%%%%%%%%%%%%%%%%

\section{Probing Black Hole Microstructures}\label{Sec:Thermodynamic Topology}

This section investigates thermodynamic geometry within the context of extended black hole thermodynamics. The enthalpy, as defined in equation (\ref{first law}), is used as the suitable thermodynamic potential to investigate the nature of microscopic interactions when Lorentz Invariance Violation (LIV) effects are present, as they appear in black holes in Lorentz-Violating gravity in AdS spacetime. The well-established analogy with van der Waals (vdW) fluids serves as the basis for this choice. We examine the effects of LIV-induced corrections on the microstructure of black holes in Lorentz-Violating gravity using Ruppeiner geometry, using enthalpy as the central potential. %We pay special emphasis to the thermodynamic state spaces that are defined by the $(S,P)$ and $(T,V)$ planes.
We examine many limiting cases arising from the modified equation of state for black holes in Lorentz-Violating gravity and investigate the type of interactions that can develop when LIV effects are introduced. Based on the thermodynamic fluctuation theory, the basic idea of this method is to express the number of microstates($\Omega$) of a thermodynamic system and entropy($S$) via the relation
\begin{equation*}\label{Omega1}
    S= k_B\;\ln{\Omega} \ ,
\end{equation*}
where, $k_B$ is the Boltzmann constant. Let's consider a thermodynamic system $I_0$ be in equilibrium with a subsystem $I$ with two independent fluctuating coordinates, $x^i$ $(i=1,2)$. Then the probability $P(x^1,x^2)$ of getting the system between $(x^1,x^2)$ and $(x^1 + dx^1,x^2+dx^2)$ states, would be related to the number of microstates. In accordance with the second law of thermodynamics, the pair $(x^1,x^2)$ chooses values that maximize the entropy $S=S_{\text{max}}$. In other words, the pair $(x^1,x^2)$ reflects thermodynamic fluctuations surrounding this maximum and expanding the entropy\footnote{It is important to note that the total entropy comprises contributions from both the system and its environment; therefore, the later can be ignored under certain conditions~\cite{Ruppeiner:1995zz}} up to second order, the probability $P(x^1, x^2)$ decays exponentially with the square of the thermodynamic length $\Delta l_R$, scaled by a factor of one-half~\cite{Ruppeiner:1995zz}. Now, we can write the line element, which quantifies the thermodynamic distance between two infinitesimally close fluctuating states as,
\begin{equation}\label{distance}
\Delta l_{R}^2 \, = -\frac{1}{k_B}\, \frac{\partial^2 S}{\partial x^i\partial x^j}\, \Delta x^i \Delta x^j \ .
\end{equation}
A smaller thermodynamic distance corresponds to a higher probability of transition between those states. Thus, the geometry encodes the relative likelihood of fluctuations: the closer two states are in this metric, the more probable the fluctuation between them. Putting emphasis on the existing results of thermodynamic curvature $R$, derived from the metric in equation~(\ref{distance}) for a variety of systems, such as liquid-gas systems, ideal and van der Waals gases, and quantum Bose/Fermi systems, an empirical understanding has been established: a negative value of $R$ is typically associated with dominant attractive interactions, while a positive value corresponds to dominant repulsive interactions in the system. A negative (positive) divergence of curvature signifies that the system is unstable (strongly coupled), which further points towards the stability of Bose (Fermi) type systems
~\cite{Wei:2019uqg, mrugala1993, Singh:2020tkf, Singh:2023hit, Singh:2023ufh, singh2022thermodynamic, Singh:2025ueu}. Specifically, the divergences of $R$ typically correspond to the critical points of the system. The condition in which there is no interaction or in which repulsive and attractive interactions are balanced is indicated by a vanishing curvature. A detailed description of the characteristic of the Ruppeiner metric in the theory of thermodynamic fluctuation can be found in~\cite{RevModPhys.67.605}. In general, the components of a metric with two fluctuation coordinates in two dimensions can be expressed as a $2\times2$ matrix, $g_{ij}$, where $i=1,2$. One can determine the Ricci scalar of the geometry defined by the Ruppeiner metric by employing the formulas for Riemannian geometry. In a thermodynamic system, the physical information of the microscopic interactions is provided by this Ricci scalar, which we will refer to as the Ruppeiner curvature. It is possible to express the Ricci scalar for this type of metric as~\cite{Singh:2020tkf},
\begin{equation} \label{R1}
R = - \frac{1}{\sqrt{g}}\bigg[\frac{\partial}{\partial x^1}\bigg(\frac{g_{12}}{g_{11}\sqrt{g}}\frac{\partial g_{11}}{\partial x^2} - \frac{1}{\sqrt{g}}\frac{\partial g_{22}}{\partial x^1}\bigg) + \frac{\partial}{\partial x^2}\bigg(\frac{2}{\sqrt{g}}\frac{\partial g_{12}}{\partial x^2} - \frac{1}{\sqrt{g}}\frac{\partial g_{11}}{\partial x^2} - \frac{g_{12}}{g_{11}\sqrt{g}}\frac{\partial g_{11}}{\partial x^1}\bigg) \bigg],
\end{equation}
where \(g=g_{11}g_{22}-g_{12}g_{21}\) is the determinant of the $2\times2$ metric tensor $g_{ij}$.

\subsection{ Ruppeiner Line Elements}
In the standard thermodynamic approach, for solutions like the Schwarzschild black hole, the formulation of a thermodynamic geometry is not viable in the absence of additional parameters characterizing the black hole, such as electric charge or angular momentum, as all thermodynamic quantities are functions of a single variable, the event horizon radius. Nevertheless, an additional degree of freedom is introduced in the framework of extended black hole thermodynamics by treating the cosmological constant as a thermodynamic variable~\cite{Henneaux:1984ji, Teitelboim:1985dp, Dolan:2010ha}. Although it demands the construction of a suitable thermodynamic line element, thereby rendering it feasible to develop a non-trivial thermodynamic geometry even for neutral black holes.

For the extended thermodynamic setup, it may be suitable to choose enthalpy as the potential, \( H = H(S, P) \), where the fluctuation coordinates are \( S \) and \( P \). In order to compute the line element of the Ruppeiner metric with independent coordinates $S$ and $P$, let us start with the generic expression,
\begin{equation}\label{dlR}
dl^2_R=-g_{\mu \nu}dx^\mu dx^\nu = -dz_\mu dx^\mu \;\;\;\; \text{where}\;\;g_{\mu \nu}=\partial \mu \partial \nu S \ .
\end{equation}
Here we have defined $dz_\mu =g_{\mu \nu}dx^\nu$. Therefore, we must have $z_\mu =\partial S/ \partial x^\mu$. By assuming entropy as a function of mass and pressure the first law of thermodynamicscan be represented as
\begin{equation}\label{firstlaw}
dS=\frac{1}{T}dM - \frac{V}{T}dP \ .
\end{equation}
using Eq.~\eqref{dlR} and Eq.~\eqref{firstlaw} and identifying $x^1 = M$ and $x^2 = P$ it is easily verified as
\begin{eqnarray}
dz_1=-\frac{dT}{T^2},\quad dz_2=\frac{V}{T^2}dT-\frac{1}{T}dV \ .
\end{eqnarray}
Now, using Eq.~\eqref{dlR}, we have
\begin{eqnarray}
dl^2_R = -dz_{1} dx^1 - dz_{2} dx^2 = \frac{dT\,dM}{T^2} -\frac{VdT\,dV}{T^2}+\frac{dP\,dV}{T} \ , \nonumber
\end{eqnarray}
Finally, utilizing the first law, the universal form of the line element can be written as 
\begin{equation}\label{Line element}
\boxed{dl^2_R=\frac{dS\,dT}{T}+\frac{dP\,dV}{T}}
\end{equation}
The Eq.~\eqref{Line element} is the function of $S, T, P$, and $V$. We choose different planes and derive the Ruppeiner line element in different planes, and irrespective of the choice of planes, the physics remains the same.
%----------
\subsubsection*{\textbf\underline{{Ruppeiner line element in the $(S,P)$-plane}}}

First we choose the $(S-P)$ plane, i.e., we choose $S$ and $P$
to be independent coordinates such that,
\begin{eqnarray}
T= T(S,P) \;\;\;\;\;\;\;;\;\;\;\;\; V= V(S,P) \nonumber
\end{eqnarray}
and consequently,
\begin{eqnarray}
d T=\bigg(\frac{\partial T}{\partial S}\bigg)_P d S + \bigg(\frac{\partial T}{\partial P}\bigg)_S d P\;\;\;\;;\;\;\;\; d V=\bigg(\frac{\partial V}{\partial S}\bigg)_P d S+\bigg(\frac{\partial V}{\partial P}\bigg)_S d P \nonumber
\end{eqnarray}
Substituting this in Eq.~\eqref{Line element} and using the Maxwell's relations
\begin{eqnarray}
   \bigg(\frac{\partial T}{\partial P}\bigg)_{S}= \bigg(\frac{\partial V}{\partial S}\bigg)_{P} \nonumber
\end{eqnarray}
we can finally express the corresponding line element in the $(S,P)$-plane as,
\begin{equation}\label{RuppeinerSP1}
dl_R^2 = \frac{1}{C_P} dS^2 + \frac{2}{T} \left( \frac{\partial T}{\partial P} \right)_S dS \, dP - \frac{V}{T B_S} dP^2 \ ,
\end{equation}
where \( B_S = -V \left( \frac{\partial P}{\partial V} \right)_S \) is the adiabatic bulk modulus.

\subsubsection*{Ruppeiner line element in the $(T,V)$-plane}

Secondly, we choose the $(T-V)$ plane, i.e., we choose $T$ and $V$ are independent coordinates, and then
\begin{eqnarray}
S=S(T,V) \;\;\;\;\;;\;\;\;\;\;  P=P(T,V) \ . \nonumber
\end{eqnarray}
Further, one gets,
\begin{eqnarray}
    d S=\bigg(\frac{\partial S}{\partial T}\bigg)_V d T+\bigg(\frac{\partial S}{\partial V}\bigg)_T d V \;\;\;\;\;;\;\;\;\;\; d P=\bigg(\frac{\partial P}{\partial T}\bigg)_V d T+\bigg(\frac{\partial P}{\partial V}\bigg)_T d V \ . \nonumber
\end{eqnarray}
Substituting this in Eq.\eqref{Line element}, the corresponding line element in the $(T, V)$-plane can be expressed as,
\begin{equation}\label{RuppeinerTv1}
  dl_R^2 = \frac{C_V}{T^2} dT^2+ \frac{2}{T} \left( \frac{\partial P}{\partial T} \right)_V dT dV + \frac{1}{T} \left( \frac{\partial P}{\partial V} \right)_T dV^2 \ ,
\end{equation}
where \( C_V \) is the specific heat at constant volume, which vanishes for static black holes. 

In AdS spacetime, entropy $S$ and thermodynamic volume $V$ are not independent variables in spherically symmetric black holes~\cite{Singh:2020tkf}. Consequently, instead of the internal energy $U(S,V)$, the enthalpy $H(S,P)$ is regarded as the fundamental thermodynamic potential. The Ruppeiner metric for black holes in Lorentz-Violating gravity is constructed in its general form using this framework, and its specific expressions are calculated in the $(S,P)$ and $(T,V)$ thermodynamic phase spaces in the later subsection below. Of course, there are many other possible line elements, depending on the thermodynamic ensemble and the nature of fluctuations. Numerous aspects of the black hole's thermodynamic behavior, including stability, phase transitions, and critical occurrences, are captured by these metrics~\cite{Wei:2015iwa,Mann2019,Wei:2019yvs,SAdS,Mirza:2007ev,Bhattacharya:2017hfj}. Importantly, we stress that all these entities are geometrically consistent: Legendre transformations mandate conformal transformations that link the thermodynamic metrics. Because each thermodynamic potential has its own set of natural variables, the accompanying metrics are conformally equivalent as one moves between them, preserving the underlying geometric structure that the Legendre framework imposes~\cite{Bravetti:2012dn,Dolan:2015xta,Ghosh:2023khd}.

\subsection{Microstructures of black holes in Lorentz-Violating gravity}
In this subsection, we consider the framework of extended black hole thermodynamics, wherein the cosmological constant $\Lambda$  is interpreted as a thermodynamic pressure $P=-\Lambda/8\pi$~\cite{Kubiznak:2012wp,Kastor:2009wy}. Within this formalism, the black hole mass $M$ is reinterpreted as the enthalpy of the system, defined by the relation $M=U+PV,$ where $U$ is the internal energy and $V$ is the thermodynamic volume. The mass can be obtained using the metric function specified in equation \eqref{Metric Linear} for the linear case and equation \eqref{Metric Quadratic} for the quadratic case.

\subsubsection{Linear Case}
The key thermodynamic quantities, namely the Hawking temperature, entropy and thermodynamic volume are, computed as
\begin{eqnarray}
T= \frac{(1+l_1) +8 \pi  P r_h^2}{4 \pi (1-l_1)r_h -2 \pi  l_2 r_h} \;\;\;\;\;;\;\;\;\;\;S=\pi r_h^2\;\;\;\;\;;\;\;\;\;\; V=\frac{4}{3}\pi r_h^3 \ .
\label{Temperature}
\end{eqnarray}
These expressions provide the foundation for analyzing the thermodynamic phase structure and geometric properties of the black hole within the extended phase space. To obtain the behavior of the thermodynamic temperature of the black hole with entropy, we plot the equation~\eqref{Temperature}.
\begin{figure}[h!]
    \begin{center}
        \includegraphics[scale=.60]{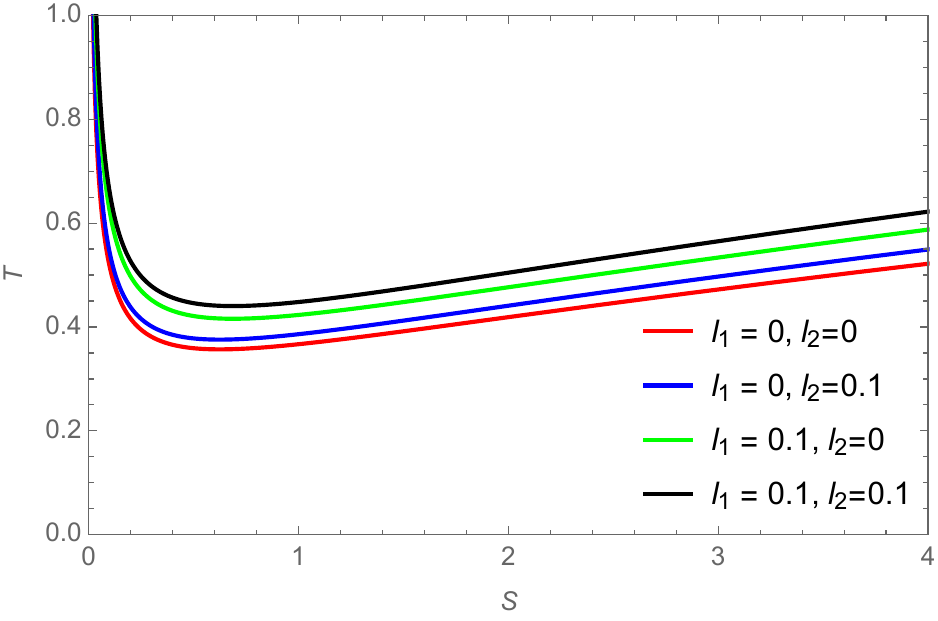} 
    \end{center}
    \caption{The behavior of black hole temperature $T$ with entropy $S$ for fixed values of $l_1, l_2$ and thermodynamic pressure $P$.}
    \label{T_vs_S}
\end{figure}
It is evident from Fig.~(\ref{T_vs_S}), that the Hawking temperature exhibits a minimum that leads to a divergence that indicates a phase transition. This divergence signifies a change from thermodynamically stable to unstable black hole structures.
%The first law of black hole thermodynamics can be expressed as, 
%\begin{equation}
%dM = TdS + VdP \ ,
%\label{First law}
%\end{equation}
Further, utilizing equation~\eqref{Temperature}, we can obtain the equation of state as
\begin{equation}
P(T, r_h) = \frac{T (1-l_1)}{2 r_h} -\frac{l_2 T}{4 r_h}-\frac{(1+l_1)}{8 \pi  r_h^2}
\label{Eos1}
\end{equation}

Utilizing equation (\ref{RuppeinerSP1}), equation (\ref{Temperature}) and equation (\ref{Eos1}), we can directly compute the Ruppeiner curvature for black holes in Lorentz-Violating gravity on the $(S,P)$-plane as,
\begin{eqnarray}
    R(S,P) =  \frac{-(1+l_1)}{S(1+ l_1 +8 P S)}
\end{eqnarray}

\begin{figure}[h!]
    \begin{center}
        \includegraphics[scale=.50]{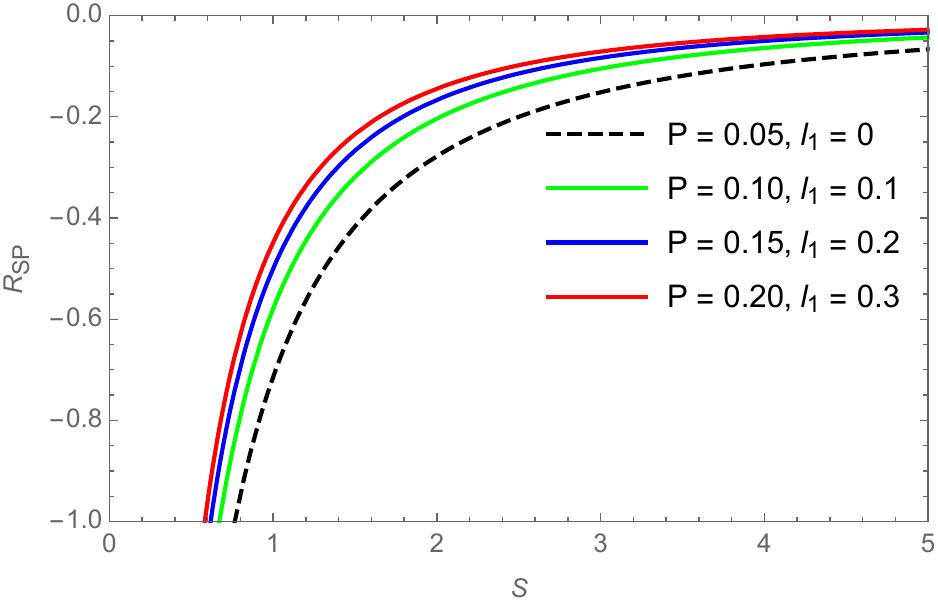}
        \includegraphics[scale=.50]{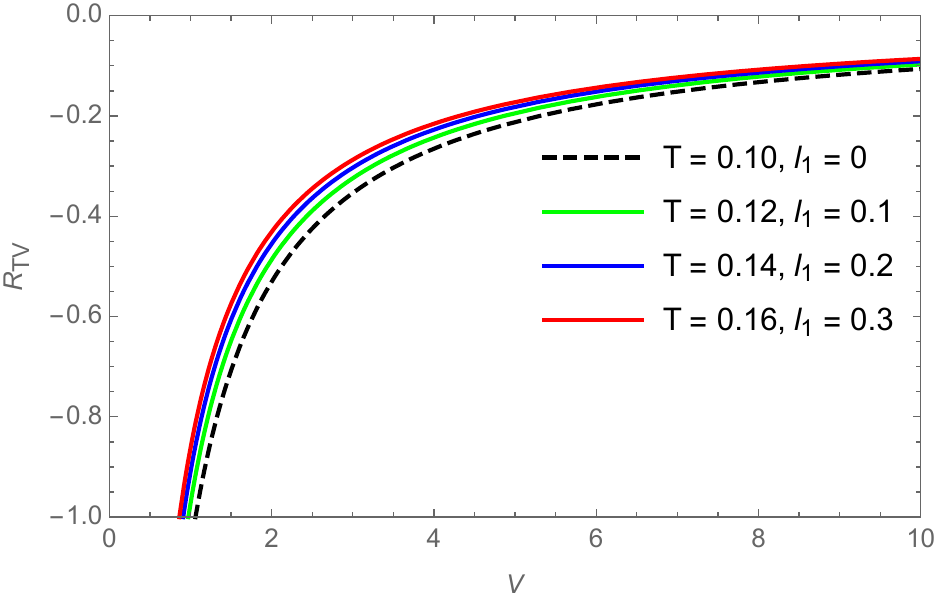}
    \end{center}
    \caption{ {\bf Left.} The behavior of the thermodynamic curvature $R_{SP}$ with the entropy $S$ of the black hole for fixed values of pressure $P$ and $l_1$ of the black holes in Lorentz-Violating gravity for the linear case. {\bf Right.} The behavior of the thermodynamic curvature $R_{TV}$ with the thermodynamic volume $V$ of the black hole for fixed values of thermodynamic temperature $T$ and $l_1$ of the black holes in Lorentz-Violating gravity for the linear case.}
    \label{Rsp_vs_S1}
\end{figure}

Again, utilizing equation (\ref{RuppeinerTv1}), equation (\ref{Temperature}) and equation (\ref{Eos1}), we can directly compute the Ruppeiner curvature for black holes in Lorentz-Violating gravity on the $(T,V)$-plane as,
\begin{eqnarray}
    R(T,V) =  \frac{-(1+l_1)}{3 \pi  TV}
\end{eqnarray}
One can note that in the limit $l_1 \rightarrow 0$, the Ruppeiner curvature matches with the curvature for Schwarzschild black holes on the  $(S,P)$ as well as $(T,V)$-plane both~\cite{SAdS}. The behavior of the thermodynamic curvature with the entropy $S$, and with thermodynamic volume $V$ respectively, for the black holes in Lorentz-Violating gravity is displayed in Fig. \ref{Rsp_vs_S1}.

\subsubsection{Quadratic case}

For the quadratic case, again the key thermodynamic quantities, namely the Hawking temperature, entropy and thermodynamic volume are, computed following exactly the same approach as,

\begin{eqnarray}
T= \frac{2 P r_h}{(1+l_1)}+\frac{(1+l_1)}{4 \pi r_h (1+3l_1)} \;\;\;\;\;;\;\;\;\;\;S=\pi r_h^2\;\;\;\;\;;\;\;\;\;\; V=\frac{4}{3}\pi r_h^3 \ .
\label{Temp2}
\end{eqnarray}
For the quadratic case also, the Hawking temperature exhibits a minimum. Again, utilizing equation~\eqref{Temp2}, we can obtain the equation of state for the quadratic case as
\begin{equation}
P(T, r_h) = \frac{\left(1+l_1\right) \left(4 \pi r_h T \left(3 l_1+1\right)-(1+l_1)\right)}{8 \pi  \left(3 l_1+1\right) r_h^2}
\label{Eos2}
\end{equation}

Utilizing equation (\ref{RuppeinerSP1}), equation (\ref{Temp2}) and equation (\ref{Eos2}), we can directly compute the Ruppeiner curvature for black holes in Lorentz-Violating gravity on the $(S,P)$-plane as,
\begin{eqnarray}
    R(S,P) = \frac{-\left(1+l_1\right){}^2}{S \left(\left(l_1+1\right){}^2 + 8 PS \left(1+3 l_1\right)\right)} 
\end{eqnarray}

%One can note that in the limit $l_1 \rightarrow 0$, the Ruppeiner curvature matches the curvature for Schwarzschild black holes on the $(S,P)$-plane given as,
%\begin{eqnarray}
%    R(S,P) =  \frac{-1}{S(1 + 8PS)}.
%\end{eqnarray}

\begin{figure}[h!]
    \begin{center}
        \includegraphics[scale=.50]{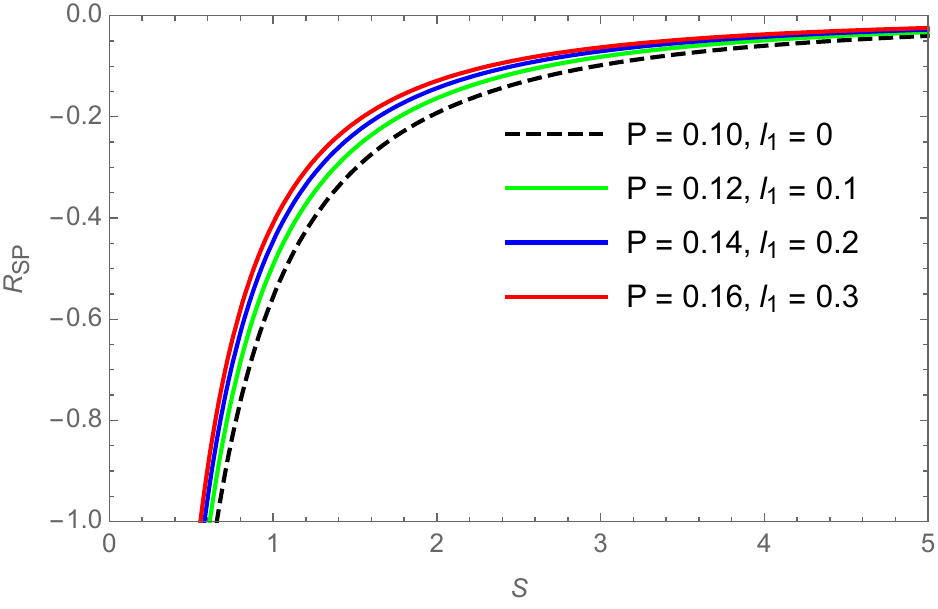}
        \includegraphics[scale=.50]{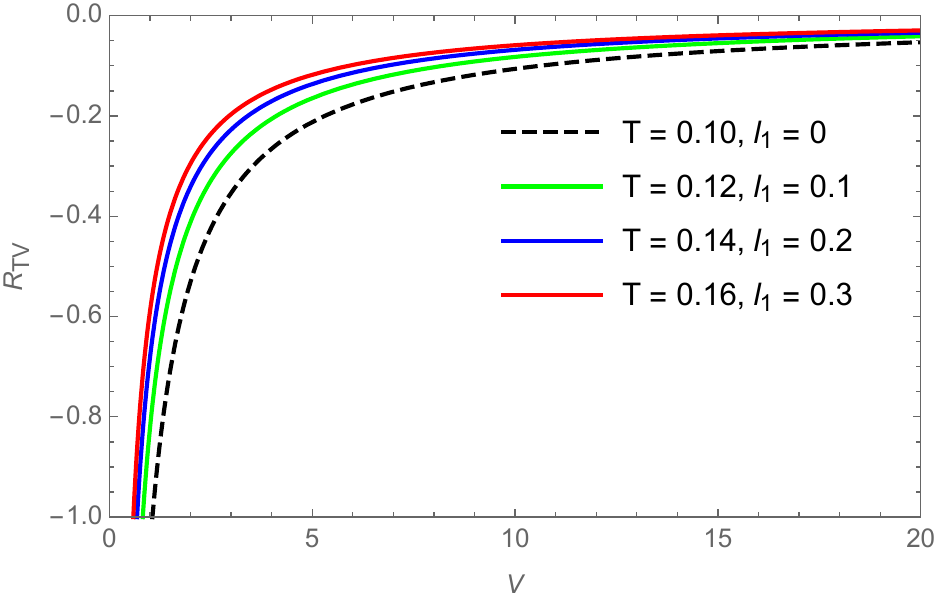}
    \end{center}
    \caption{ {\bf Left.} The behavior of the thermodynamic curvature $R_{SP}$ with the entropy $S$ of the black hole for fixed values of pressure $P$ and $l_1$ of the black holes in Lorentz-Violating gravity for the quadratic case. {\bf Right.} The behavior of the thermodynamic curvature $R_{TV}$ with the thermodynamic volume $V$ of the black hole for fixed values of thermodynamic temperature $T$ and $l_1$ of the black holes in Lorentz-Violating gravity for the quadratic case.}
    \label{Rsp_vs_S2}
\end{figure}

Further, utilizing equation (\ref{RuppeinerTv1}), equation (\ref{Temp2}) and equation (\ref{Eos2}), we can directly compute the Ruppeiner curvature for black holes in Lorentz-Violating gravity on the $(T,V)$-plane as,
\begin{eqnarray}
    R(T,V) = -\frac{1+2 l_1 + l_1^{2}}{3 \pi TV(1+3l_1)}
\end{eqnarray}
One can again note that, in the limit $l_1 \rightarrow 0$, the Ruppeiner curvature matches the curvature for Schwarzschild black holes on the $(S,P)$ as well as $(T,V)$-planes, both, for the quadratic case~\cite{SAdS}.
%One can again note that in the limit $l_1 \rightarrow 0$, the Ruppeiner curvature matches the curvature for Schwarzschild black holes on the $(T,V)$-plane given as,
%\begin{eqnarray}
%    R(T,V) = \frac{-1}{3 \pi  TV}
%\end{eqnarray}
The behavior of the thermodynamic curvature for the black holes in Lorentz-Violating gravity for the quadratic case is displayed in Fig. \ref{Rsp_vs_S2}.

Within the framework of extended thermodynamics, to investigate the statistical interactions of black hole microstructures, the Ruppeiner geometry serves as an important tool for the Kalb–Ramond black holes incorporating Lorentz-invariance violation. It is evident that the Ruppeiner scalar curvature $R$, computed in both the $(S,P)$ and $(T,V)$ thermodynamic planes, consistently exhibits negative values. It should be noted that both $R_{SP}$ and $R_{TV}$ asymptotically diverge as the black hole becomes extremal, i.e. $T=0$~\cite{Singh:2020tkf}. In addition, the Ruppeiner curvatures derived in both planes are equivalent and we will denote the Ruppeiner curvature as $R$ later on. Furthermore, this persistent negativity of $R$ clearly indicates that the dominant interaction among microstructures of the black holes in Lorentz-Violating gravity is attractive. Such attractive interactions are a hallmark of systems with correlated or clustered microstate, akin to the repulsive systems like ideal gases where $R \approx 0$ or is positive.

\quad Moreover, the coupling parameters associated with the Kalb–Ramond field significantly modifies the magnitude of the Ruppeiner curvature. With an increase in the coupling parameter $l_1$, $R$ becomes larger, implying that the strength of attractive interactions becomes enhanced. This implies that, especially in regimes of small entropy or volume, the coupling parameters strengthen the effective microscopic binding or correlation among the degrees of freedom of black holes. It is interesting to note that the Ruppeiner curvature remains finite and smooth during a phase transition, even while the specific heat $C_P$ diverges. This behavior arises from the curvature itself, the divergence in $C_P$ does not induce a corresponding divergence in $R$ because the critical singularity is suppressed by a vanishing prefactor. In addition to indicating that the microstructure is attraction-dominated, the Ruppeiner geometry for black holes in Lorentz-Violating gravity demonstrates that the Lorentz-invariance violation parameters enhance these attractive interactions without introducing curvature singularities at the phase transition. Remarkably, in the limit $l_1 \rightarrow 0$ and $l_2 \rightarrow 0$, the Kalb–Ramond black hole reduces exactly to the Schwarzschild case, indicating that any deviation in thermodynamic behavior occurs entirely from the presence of coupling parameters. The Schwarzschild-AdS black hole has a negative and finite Ruppeiner scalar curvature in the $(S,P)$ and $(T,V)$ planes, suggesting weakly attractive microstructure interactions, despite the fact that such neutral black holes do not admit a thermodynamic geometry in the traditional phase space. It is apparent that the thermodynamic geometry is robust and non-singular in this regime primarily because the Ruppeiner curvature in both planes remains smooth and finite, near entropy, where the specific heat diverges. Consequently, the KR black hole, through Lorentz violation effects, shows a system with significantly enhanced but nonetheless attractive interaction, whereas the Schwarzschild-AdS black hole provides a baseline state for weakly interacting microstructures. This behavior affirms the role of Lorentz violation in enriching the microphysical landscape of black holes in a thermodynamically consistent and geometrically stable manner.

\section{Geodesics in Lorentz-Violating gravity spacetime}\label{Sec:Shadow}

In order to quantify how Lorentz-violating effects alter the trajectories of both lightlike and timelike probes, we first review geodesic motion in the black-hole spacetime. Following Ref.\cite{Duan:2023gng}, the world-line of a freely falling particle extremises the action
\begin{equation}
    S = \int \! \mathcal{L}\,\mathrm d\lambda, 
    \qquad 
    \mathcal{L} = -\tfrac12\, g_{\alpha\beta}\,\dot{x}^{\alpha}\dot{x}^{\beta},
\end{equation}
where an over-dot denotes differentiation with respect to the affine parameter~$\lambda$.
Upon using the normalisation condition for the four-velocity we may rewrite the Lagrangian as
\[
   \mathcal{L} = \frac{\varepsilon}{2},
   \qquad\varepsilon =
   \begin{cases}
      0 & \text{for photons},\\
      1 & \text{for massive particles}.
   \end{cases}
\]
Because the metric of interest is spherically symmetric and static,
\begin{equation*}
  \mathrm d s^{2} 
    = -f(r)\,\mathrm d t^{2}+\frac{\mathrm d r^{2}}{f(r)}
      +r^{2}(\mathrm d\theta^{2}+\sin^{2}\theta\,\mathrm d\phi^{2}),
\end{equation*}
one can, without loss of generality, confine the motion to the equatorial plane
$\theta=\pi/2$.  Substituting the metric into $\mathcal{L}$ then yields
\begin{equation}
    f(r)\!\left(\frac{\mathrm d t}{\mathrm d\lambda}\right)^{\!2}
    - \frac{1}{f(r)}\!\left(\frac{\mathrm d r}{\mathrm d\lambda}\right)^{\!2}
    - r^{2}\!\left(\frac{\mathrm d\phi}{\mathrm d\lambda}\right)^{\!2}
    = \varepsilon.
    \label{eq:lagrangian-equation}
\end{equation}
Stationarity ($\partial_t$ Killing vector) and axial symmetry ($\partial_\phi$ Killing vector) guarantee two constants of motion:
\begin{align}
   E &= \frac{\partial\mathcal L}{\partial\dot t}
      = f(r)\,\frac{\mathrm d t}{\mathrm d\lambda},
      &&\text{specific energy}, \label{EE}\\
   L &= -\frac{\partial\mathcal L}{\partial\dot\phi}
      = r^{2}\,\frac{\mathrm d\phi}{\mathrm d\lambda},
      &&\text{specific angular momentum}.\label{LL}
\end{align}
Eliminating $\dot t$ and $\dot\phi$ in favour of $E$ and $L$ enables us to
cast the radial equation \eqref{eq:lagrangian-equation} into the form
\begin{equation}
   \dot r^{2} 
   = E^{2} - f(r)\!\left(\varepsilon^{2} + \frac{L^{2}}{r^{2}}\right)
   = E^{2} - V_{\mathrm{eff}}^{2}(r),
   \label{eq:radial-motion}
\end{equation}
where we have introduced the \emph{effective potential}
\begin{equation}
   V_{\mathrm{eff}}(r)
   = \sqrt{\,f(r)\!
            \left(\varepsilon^{2} + \frac{L^{2}}{r^{2}}\right)}.
\end{equation}
Graphing $V_{\mathrm{eff}}(r)$ immediately reveals the nature of possible
orbits:
\begin{figure}
\includegraphics[width=8.5 cm]{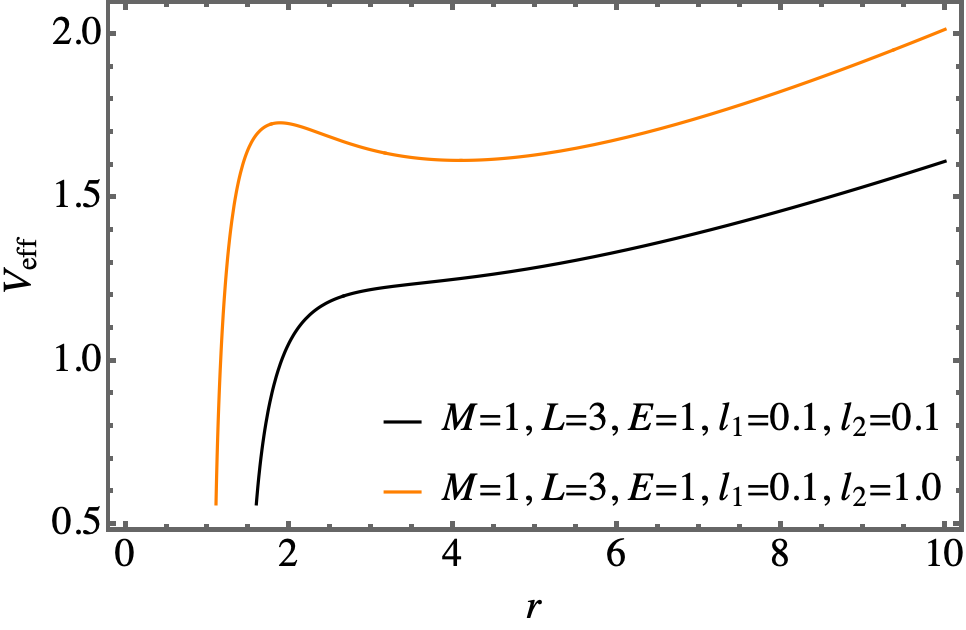}
\includegraphics[width=8.5 cm]{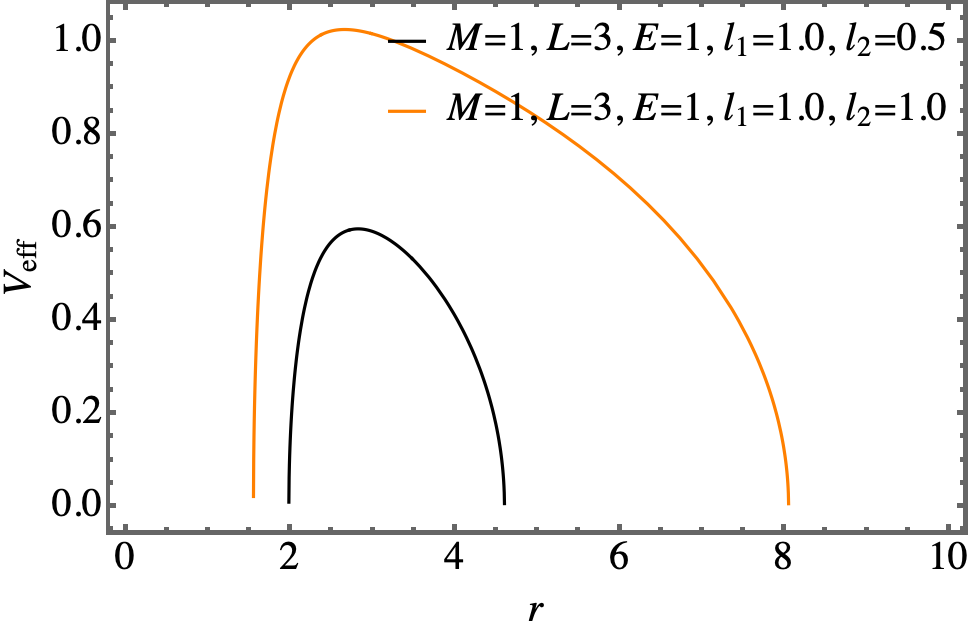}
\caption{Plots of the effective potential of particles moving around a black holes in Lorentz-violating gravity. Here we have consider $f_{1}(r)$ and kept $M=1,\,L=3.0,\,E=1.0,\,l_{1}=0.1$ fixed with $l_{2}=0.1\,\&\,1.0$ (Left panel); and $l_{2}=0.5\,\&\,1.0$ (Right panel).}\label{veff1}
\end{figure}
\begin{itemize}
  \item Roots of $E^{2}=V_{\mathrm{eff}}^{2}(r)$ correspond to turning points.
  \item Minima (maxima) of $V_{\mathrm{eff}}$ signal stable (unstable) circular orbits.
  \item For photons ($\varepsilon=0$) the peak of $V_{\mathrm{eff}}$ encodes the photon sphere, which delineates the black-hole shadow.
\end{itemize}
\begin{figure}
\includegraphics[width=5.5 cm]{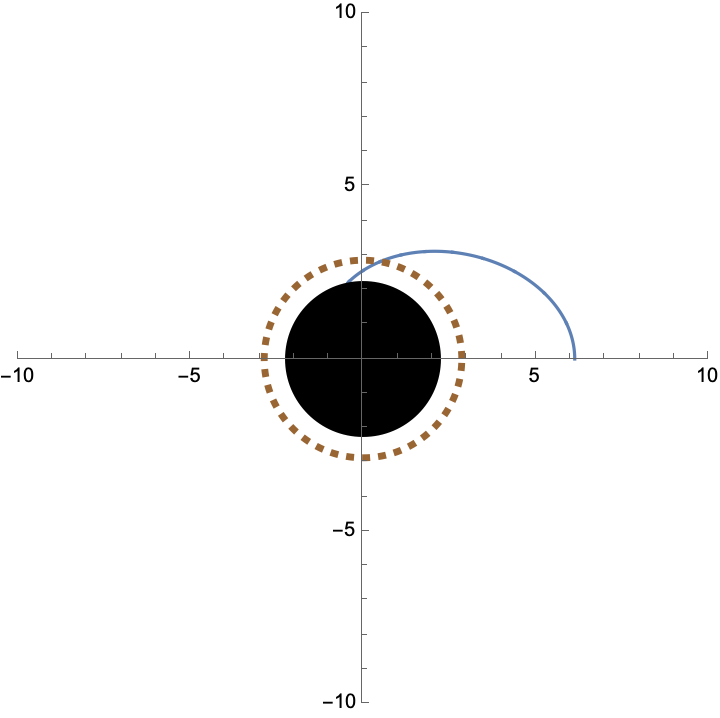}
\includegraphics[width=5.5 cm]{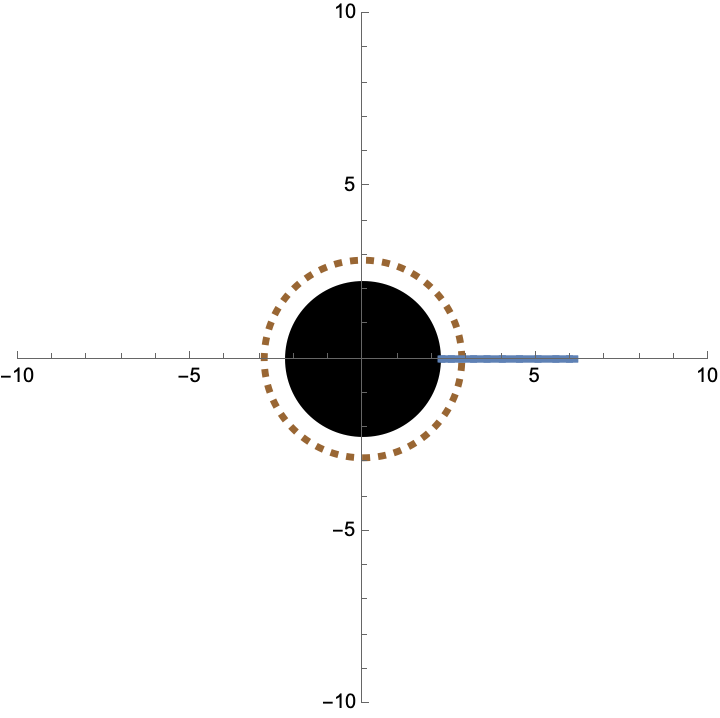}
\includegraphics[width=5.5 cm]{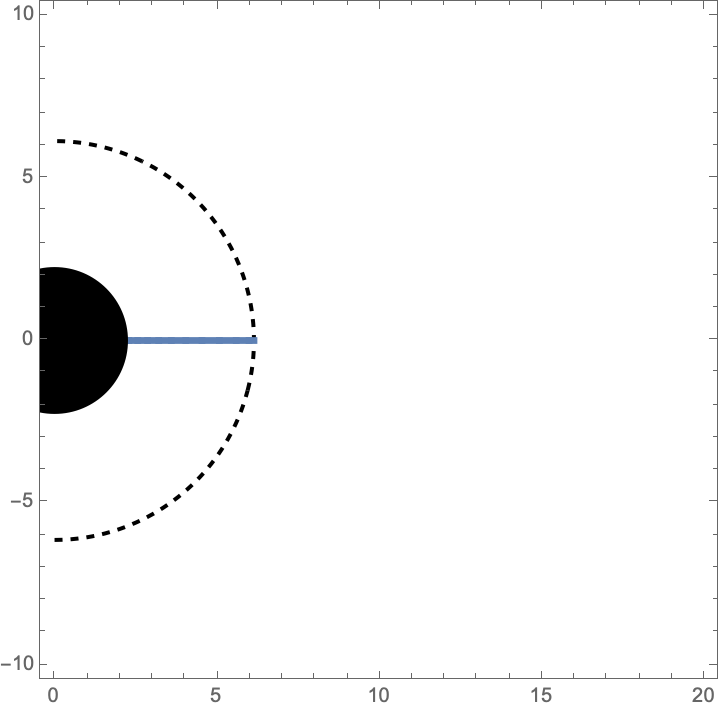}
\includegraphics[width=5.5 cm]{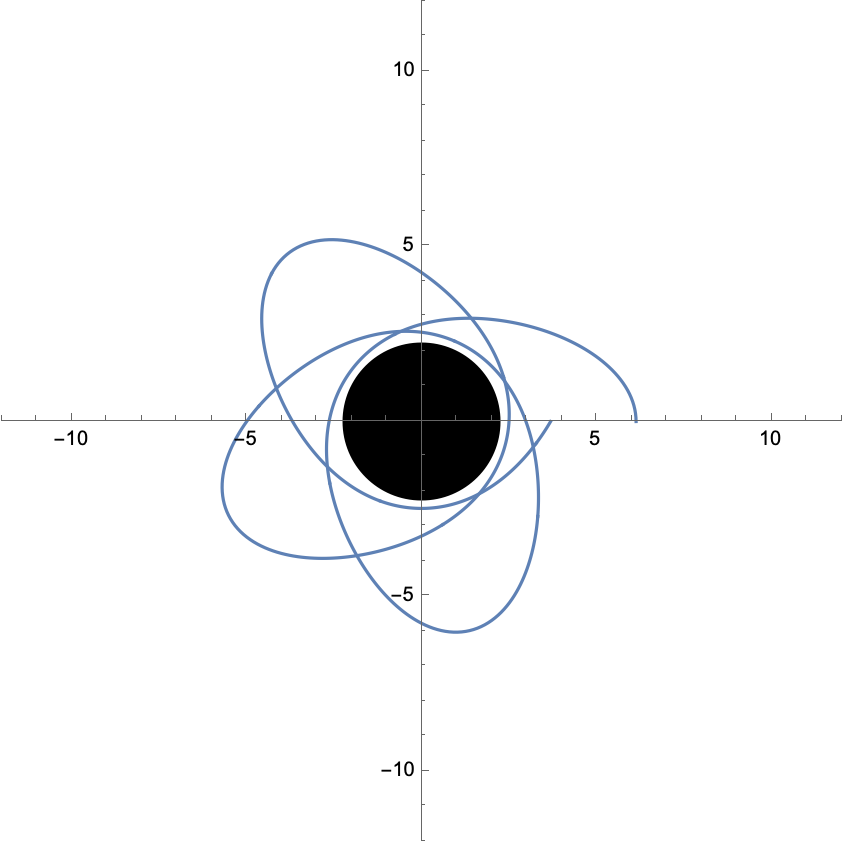}
\includegraphics[width=5.5 cm]{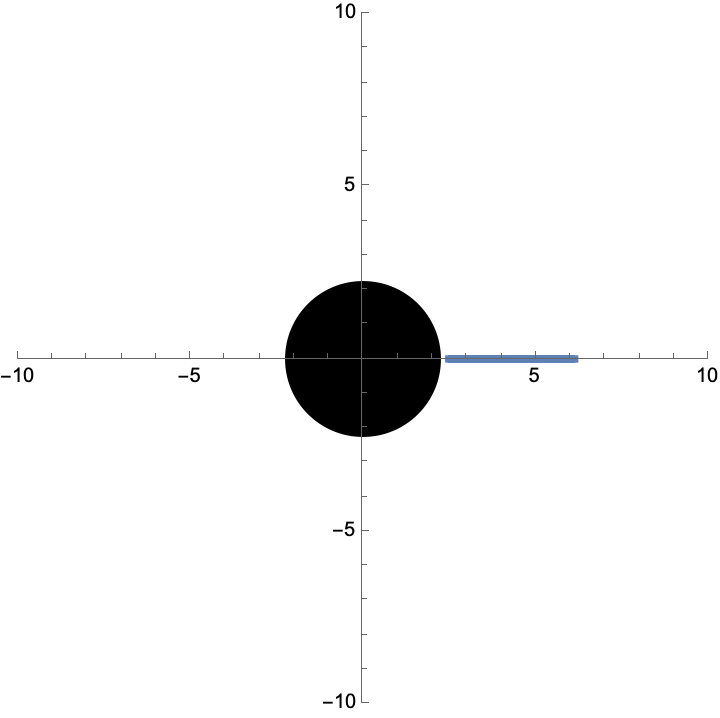}
\includegraphics[width=5.5 cm]{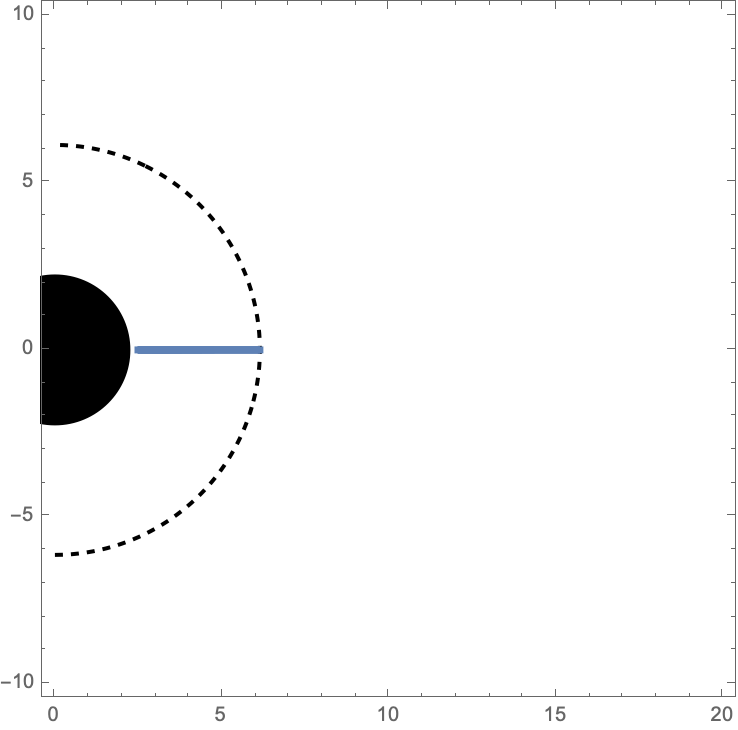}
\caption{Typical trajectories of test particles (solid curves) orbiting a black holes in Lorentz-violating
gravity (solid circle), giving also $x-y$ and
$x-z$ sections (first and second column) for $f_{1}(r)$. Due to the conservation of the particle specific energy $E$ and specific angular momentum $L$, the trajectory in 4D configuration space $(t, x, y, z)$ can be represented in
2D $x-z$ graph (third column), where we also plotted the boundary of the particle motion given by the eﬀective potential (black-dashed
curve). The photon sphere is represented by a brown-dashed
curve. We have used for the first row: $M = 1.0,\,\epsilon = 1.0,\,L= 3.0,\,E = 1.0,\,l_{1}= 0.1,\, l_{2}= 0.1,\,\Lambda = -0.0251327$, and the second one: $M = 1.0,\,\epsilon = 1.0,\,L= 3.0,\,E = 1.0,\,l_{1}= 0.1,\, l_{2}= 1.0,\,\Lambda = -0.0251327$.}\label{fm}
\end{figure}
Among all possible trajectories, \emph{stable circular orbits} play a
distinguished role: a particle can remain on such an orbit if the radial
effective potential attains a local minimum.  Formally, let
$V_{\mathrm{eff}}(r)$ denote the effective potential and $E$ the specific
energy of the particle.  Then stable circular motion requires \cite{Yang:2021chw}:
\begin{equation}
  V_{\mathrm{eff}} = E,\qquad
  \frac{dV_{\mathrm{eff}}}{dr} = 0,\qquad
  \frac{d^{2}V_{\mathrm{eff}}}{dr^{2}} > 0.
\end{equation}
The \emph{innermost stable circular orbit} (ISCO) corresponds to the
boundary case where the potential's curvature vanishes,
$d^{2}V_{\mathrm{eff}}/dr^{2}=0$.  In the Novikov–Thorne thin-disk
picture, the ISCO marks the disk's inner edge; for inspiralling compact
binaries, it signals the transition to a plunge, profoundly shaping the
gravitational-wave signal.  Consequently, any modification stemming from
Lorentz-violating effects could manifest in forthcoming observations. For timelike geodesics we set $\varepsilon = 1$ in the radial equation
of motion
\begin{equation}
  \dot r^{2}
  = E^{2} - f(r)\!\left(1+\frac{L^{2}}{r^{2}}\right)
  = E^{2} - V_{\mathrm{eff}}^{2},
  \label{eq:radial_timelike}
\end{equation}
where $L$ is the specific angular momentum and the effective potential
reads
\begin{equation}
  V_{\mathrm{eff}}(r)
  = \sqrt{f(r)\!\left(1+\frac{L^{2}}{r^{2}}\right)}.
\end{equation}
Imposing the circular-orbit conditions $V_{\mathrm{eff}}=E$ and
$dV_{\mathrm{eff}}/dr=0$ yields
\begin{align}
E^{2} &=
\frac{2\,f(r)^{2}}{2f(r)-rf'(r)},
\label{eq:E2_circ}
\\[4pt]
L^{2} &=
\frac{r^{3}f'(r)}{2f(r)-rf'(r)}.
\label{eq:L2_circ}
\end{align}
Here $f(r)$ is the lapse function given in Eq.~(\ref{Metric Linear}) and (\ref{Metric Quadratic}), $\ell_{1},\,\ell_{2}$ denote the Lorentz-violating parameters, and $M$,\,$\Lambda$ carry their usual meanings. Primes denote derivatives with respect to $r$. To study the orbits, we make the change of variable $u = 1/r$ and obtain
\begin{equation}
\left( \frac{du}{d\phi} \right)^2 = \frac{E^2}{L^2} - f(r) \left( \frac{\epsilon}{L^2} + u^2 \right) \equiv g(u).
\end{equation}
We can calculate the photon sphere radius, $r_{ph}$. For the null geodesics the radius of $r_{ph}$ can be determined
by the angular momentum’s minimum value, $L = L(r)$, which
is obtained by solving $V_{\text{eff}}(r)=0$. We display the effective potential for $f_{1}(r)$ and $f_{2}(r)$ in Figs.(\ref{veff1}) and (\ref{veff2}).

%------------------------------------------
\subsection{Linear Case}
In this subsection, we study the trajectories of timelike test particles in the Kalb-Ramond black hole spacetime governed by the linear metric function \( f_1(r) \). Our results display in Fig.~(\ref{fm}). The figure compares the particle dynamics under two different strengths of the pseudo-magnetic Lorentz-violating coupling parameter \(\ell_2\), while keeping the mass \(M = 1\) and the gravitational Lorentz-violating parameter \(\ell_1 = 0.1\) fixed. The analysis considers particles with energy \(E = 1\) and angular momentum \(L = 3\), representative of bound or plunging geodesics. In the top row (\(\ell_2 = 0.1\)), the effective potential allows for a bound orbit characterized by two turning points. The particle exhibits a stable, precessing orbit outside the photon sphere, indicating that the gravitational potential is sufficiently shallow to prevent immediate capture. The moderate precession rate per orbit reflects the curvature effects induced by the Lorentz-violating background. In contrast, the bottom row (\(\ell_2 = 1.0\)) corresponds to a stronger Lorentz-violating coupling. This significantly alters the spacetime geometry: the potential well becomes steeper and deeper, reducing the outer turning point and shifting the photon sphere closer to the event horizon. As a result, the particle quickly plunges inward after crossing the photon-sphere radius, with no outer turning point to stabilize the orbit. This scenario leads to rapid inspiral and eventual capture by the black hole. Additionally, the orbital precession is enhanced, reflecting the increased curvature and stronger gravitational field near the central object. These observations demonstrate that the pseudo-magnetic coupling \(\ell_2\) has a profound effect on the motion of particles and the structure of the effective potential. A larger \(\ell_2\) compresses the potential well and alters key orbital features such as the location of the photon sphere, the stability of circular orbits, and the overall dynamics of infalling matter. These modifications could have direct observational consequences, notably on the size and shape of the black hole shadow, the dynamics of accretion flows, and the gravitational wave signals from inspiraling objects. Thus, Figure~8 highlights how Lorentz-violating effects encoded in the Kalb-Ramond framework manifest in the astrophysical behavior of matter around black holes.
%-----------------------------------

\subsection{Quadratic Case}
In this subsection, we illustrates the typical trajectories of test particles orbiting a black hole in Lorentz-violating gravity, modeled through a nonminimally coupled Kalb--Ramond field given by $f_{2}(r)$. The plots (\ref{fm2}) provide three different perspectives: the $x\text{-}y$ and $x\text{-}z$ projections (first and second columns, respectively), and a 2D representation in the $x\text{-}z$ plane (third column), where the effective potential boundary (black-dashed curve) and photon sphere (brown-dashed curve) are also depicted. The orbits shown correspond to conserved particle energy $E = 1.0$ and angular momentum $L = 3.0$, with mass $M = 1.0$ and cosmological constant $\Lambda = -0.0251327$. Two cases are presented for comparison: $\ell_1 = 0.1$ and $\ell_1 = 0.5$, with $\ell_1$ representing the effective coupling parameter arising from Lorentz-violating modifications. As $\ell_1$ increases, the shape and spatial extent of the particle trajectories are significantly altered. For lower $\ell_1$, the orbits extend farther in the spatial directions, whereas higher values of $\ell_1$ induce a more confined trajectory, suggesting a deeper effective potential well. This behavior reflects the increasing influence of Lorentz-violating effects on the local curvature of spacetime. The presence of a photon sphere, marked by the brown-dashed curve, provides an important boundary between bound particle motion and regions of strong light deflection. While the test particles are massive, their trajectories approach but remain bounded within the effective potential, indicating stability under the chosen parameters.
\begin{figure}
\includegraphics[width=8.5 cm]{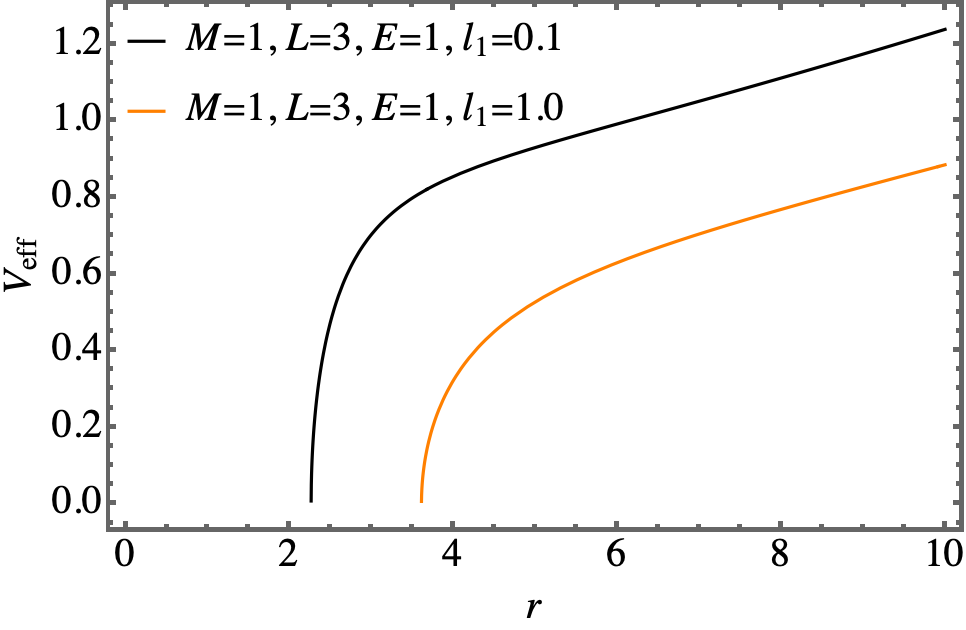}
\includegraphics[width=8.5 cm]{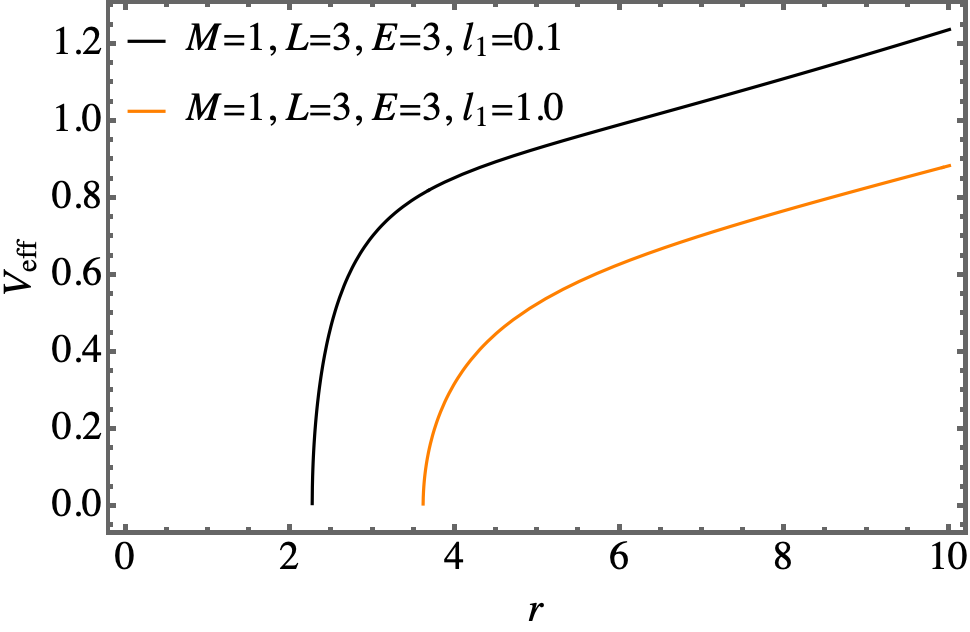}
\caption{Plots of the effective potential of particles moving around a black holes in Lorentz-violating gravity. Here we have consider $f_{2}(r)$ and used $M=1,\,L=3.0,\,E=1.0$ fixed with $l_{1}=0.1\,\&\,1.0$ (Left panel); and $M=1,\,L=3.0,\,E=3.0$ fixed with $l_{1}=0.1\,\&\,1.0$ (Right panel)}\label{veff2}
\end{figure}
\begin{figure}
\includegraphics[width=5.5 cm]{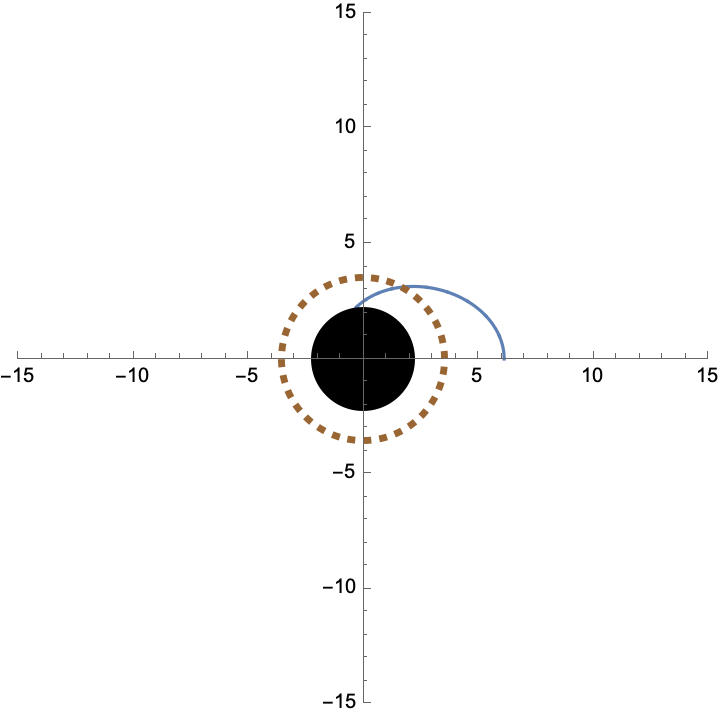}
\includegraphics[width=5.5 cm]{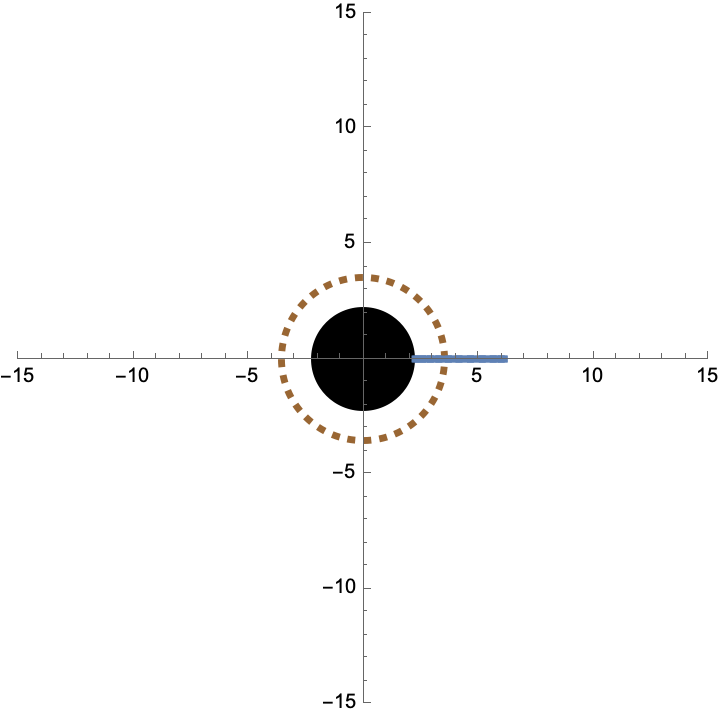}
\includegraphics[width=5.5 cm]{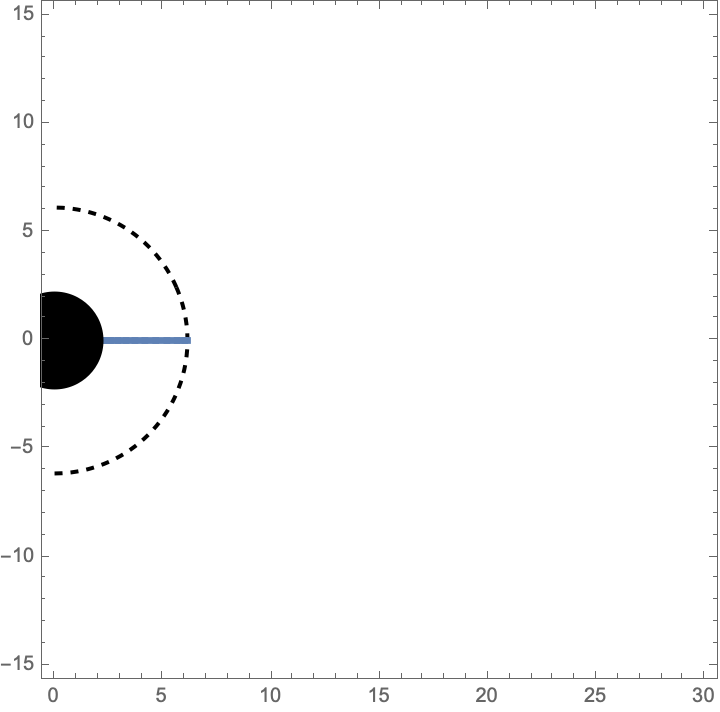}
\includegraphics[width=5.5 cm]{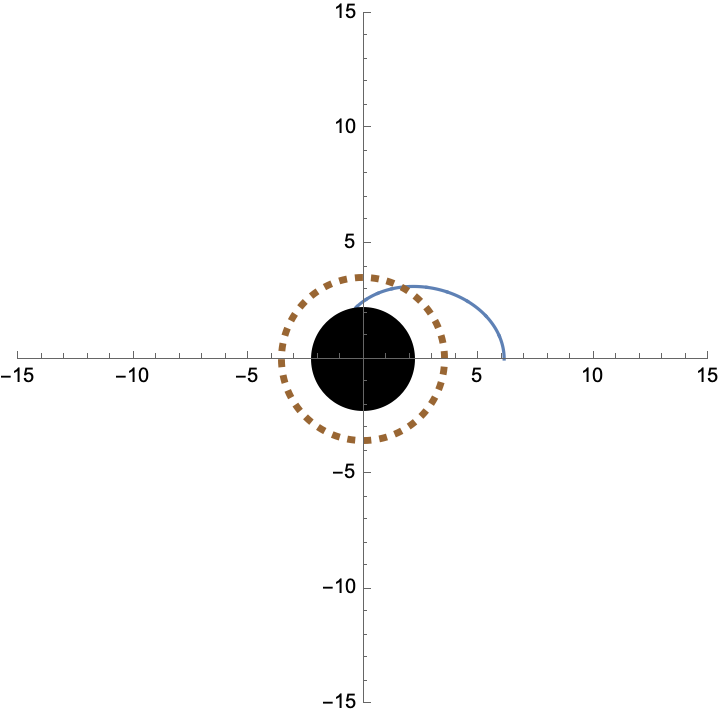}
\includegraphics[width=5.5 cm]{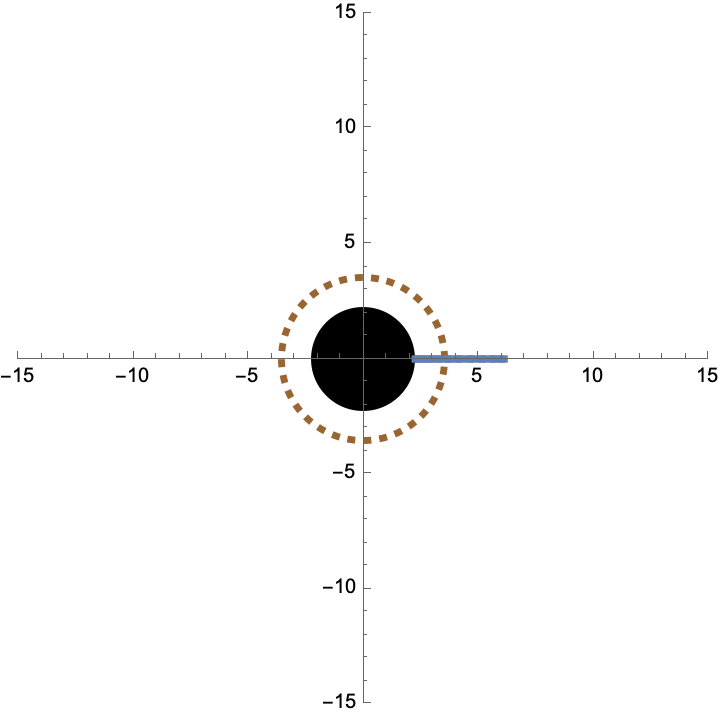}
\includegraphics[width=5.5 cm]{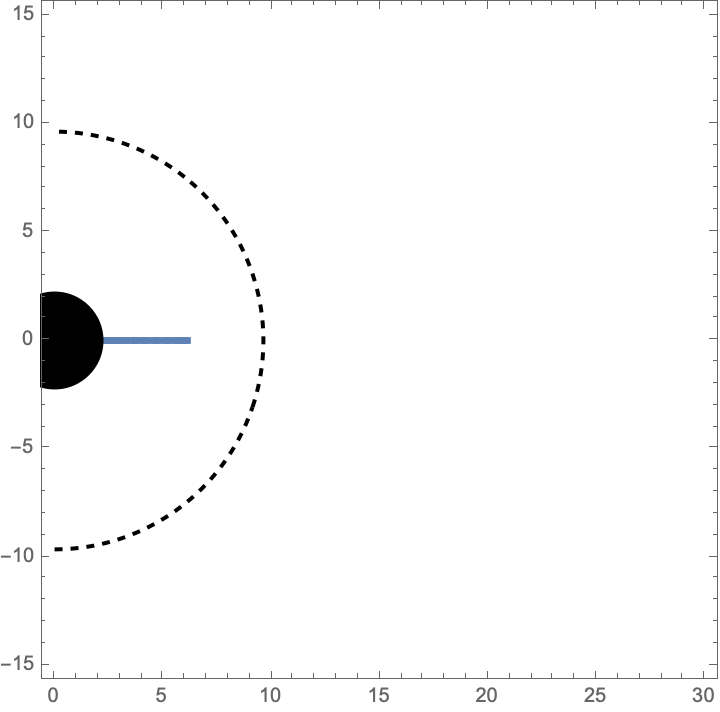}
\caption{Typical trajectories of test particles (solid curves) orbiting a black holes in Lorentz-violating
gravity (solid circle), giving also $x-y$ and
$x-z$ sections (first and second column) for $f_{2}(r)$. Due to the conservation of the particle specific energy $E$ and specific angular momentum $L$, the trajectory in 4D configuration space $(t, x, y, z)$ can be represented in
2D $x-z$ graph (third column), where we also plotted the boundary of the particle motion given by the eﬀective potential (black-dashed
curve). The photon sphere is represented by a brown-dashed
curve. We have used for the first row: $M = 1.0,\,\epsilon = 1.0,\,L= 3.0,\,E = 1.0,\,l_{1}= 0.1,\,\Lambda = -0.0251327$, and the second one: $M = 1.0,\,\epsilon = 1.0,\,L= 3.0,\,E = 1.0,\,l_{1}= 0.5,\,\Lambda = -0.0251327$.}\label{fm2}
\end{figure}
 
%%%%%%%%%%%%%%%%%%%%%%%%%%%%%%%%%%%%%%%%%%%%%%%%%%%%%%%%%%%

%%%%%%%%%%%%%%%%%%%%%%%%%%%%%%%%%%%%%%%%%%%%%%%%%%%%%%%%%%%
\section{Conclusion}\label{Sec:Conclusion}
This paper studied the thermodynamics and the universality of black holes in Lorentz-violating gravity. Its thermodynamic topology helps to compute the topological
charges and the thermodynamic geometry of these black holes. We have studied the motion of timelike test particles in these black hole spacetimes by analyzing the effective potential shaped by the Lorentz-violating couplings. The resulting dynamics reveal the existence of bound orbits and stable circular trajectories, with the location of the innermost stable circular orbit and turning points significantly influenced by the parameters $\ell_{1,2}$, and the cosmological constant.

The thermodynamics and its universality are motivated by the insight that modifications in the perturbative parameter $\epsilon$, inherent to the underlying spacetime geometry, induce coupled variations in both the horizon structure and thermodynamic characteristics of black holes. Such a dynamical interplay not only supports the essential conclusions drawn in \cite{Goon:2019faz, Anand:2025btp}, but also extends them to more generalized and complex configurations: (i) scenarios where black holes reside in Lorentz-violating backgrounds parametrized by $\epsilon$, and (ii) systems in which the mass $M$ of the black hole depends on a collection of thermodynamic parameters such as entropy $S$, pressure $P$, and geometric scales like $ \ell_1, \ell_2$, etc. The strength of this approach lies in its computational elegance and general applicability. By establishing proportional relationships between the perturbed black hole mass and other thermodynamic variables, we provide a framework that lends itself naturally to the investigation of the Weak Gravity Conjecture (WGC). Given the WGC’s pivotal role in quantum gravity, this framework offers a promising avenue for identifying viable consistency conditions for low-energy effective theories. This study is twofold: First, the derivation of a generalized universal relation expressing the black hole energy as a function of multiple thermodynamic degrees of freedom; and second, the formulation of an extended Gauss–Planck-type relation capturing the dependence of each thermodynamic quantity on higher-order perturbative corrections via the parameter $\epsilon$. Collectively, these results contribute meaningfully to the refinement of theoretical tools in black hole thermodynamics and offer deeper perspectives on gravitational consistency criteria within quantum frameworks.

After this, we have performed a thorough investigation of the thermodynamic topology of black holes in Lorentz-violating gravity, focusing on the distribution and classification of topological charges. Our analysis, supported by normalized vector field visualizations, reveals two principal topological configurations: one exhibiting a single zero point and another featuring two distinct zero points in the $(r_h, \Theta)$ parameter space. These zero points correspond to localized topological charges whose presence and structure persist under variations of free parameters, yielding total topological charges of either $W = 0$ or $W = -1$. This robust classification is further substantiated through winding number calculations, confirming the stability characteristics of these black hole solutions. Extending this framework, we mapped the free energy landscape as a scalar field whose extremum points align with the vector field zeros, thereby offering a geometric method to assign topological charges. Comparative analysis with classical black hole models—Schwarzschild and Reissner-Nordström black holes—confirms consistency with known topological charge values and highlights the universality of this topological approach in black hole thermodynamics.

A significant outcome of our study is the identification and characterization of photon spheres as topological defects within the spacetime geometry. The total topological charge of photon spheres remains invariant at $PS = -1$ across a broad range of parameter variations, reflecting their intrinsic instability and underscoring their role as critical features in gravitational systems. This topological perspective provides a powerful diagnostic tool for understanding the stability and dynamical properties of photon spheres and their interplay with black hole thermodynamics. Our results further illuminate the relationship between photon sphere topology, thermodynamic charges, and gravitational optics phenomena such as lensing and shadow formation. The systematic correlation between positive winding numbers and thermodynamic stability offers a refined criterion for classifying black hole phases and transitions. Overall, this study advances the topological understanding of black hole solutions and paves the way for future explorations into their geometric, thermodynamic, and astrophysical implications.

\quad One particularly remarkable and resilient feature of black holes in Lorentz-Violating gravity in extended phase space is that the Ruppeiner curvature is universally attractive across various thermodynamic planes. This universal nature refers to the fact that the Ruppeiner curvature, computed in different thermodynamic coordinate representations such as $(S,P)$, $(T,P)$, or $(T,V)$, consistently maintains a negative sign throughout the allowed physical domain. Regardless of the particular thermodynamic ensemble, the negative curvature consistently indicates that the underlying microstructures have dominant attractive interactions~\cite{Wei:2015iwa, Mann2019, Wei:2019yvs}. This behavior is inherent to the Kalb-Ramond system and is consistent across all widely used thermodynamic state spaces, rather than being a consequence of coordinate artifacts. This demonstrates a profound thermodynamic universality; the geometric signature of the black hole microstructure is attraction-dominated, irrespective of the choice of potential or natural variables. The sign of the Ruppeiner curvature may change across parameter space in different black hole solutions, particularly those with electric charge or scalar fields, indicating regions with repulsive and attractive interactions or microphysical behavior switching caused by criticality. As a consequence, the black holes in Lorentz-Violating gravity in the extended phase space are a key example of a geometrically consistent and thermodynamically universal system in which the attraction among the microstructures remains intact over a range of coordinate choices. Additionally, this universality suggests a sort of ensemble independence of the microscopic interpretation; under Legendre transformations, the effective interaction behavior inferred from thermodynamic geometry remains invariant. Without compromising this universality, the effect of Lorentz invariance violation enhances these attractive interactions even more, providing an intriguing window into how string-inspired or field-theoretic extensions modify the black hole microstructure without changing its qualitative thermodynamic behavior.

\quad Rather than just a coordinate artifact, this universality across thermodynamic ensembles is a fundamental characteristic of the Kalb–Ramond black hole systems. The qualitative microphysical behavior is preserved while the interaction strength is boosted by the LIV-induced coupling parameters. With regard to extended thermodynamics, black holes in Lorentz-violating gravity are therefore a reliable and consistent model that provides information on how LIV can enhance black hole microphysics without compromising geometric or thermodynamic consistency. Thus, our result demonstrates the universality and stability of the thermodynamic geometry of black holes in Lorentz-violating gravity.

%%%%%%%%%%%%%%%%%%%%%%%%%%%%%%%%%%%%%%%%%%%%%%%%%%%%%%%%%%%

\section*{Acknowledgements}
A.A. is financially supported by the Institute's postdoctoral fellowship at IITK. A.S. would like to thank CSIR-HRDG for the financial support received as a Postdoctoral Research Associate working under Project No. 03WS(003)/2023-24/EMR-II/ASPIRE.

%%%%%%%%%%%%%%%%%%

%%%%%%%%%%%%%%%%%%%%%%%%%%%%%%%%%

%%%%%%%%%%%%%%%%%%%%%%%%%%%%%%%%%%%%%%%%%%%%%%%%%%%%%%%%%%%%%%%%%%%%%%%%%%%%%%%%%%

\bibliography{KR}

\end{document}